\documentclass[conference]{IEEEtran}
\IEEEoverridecommandlockouts
\usepackage{cite}
\usepackage[font=footnotesize,caption=false,subrefformat=parens,labelformat=parens]{subfig}
\usepackage{amsmath,amssymb,amsfonts}
\usepackage{algorithmic}
\usepackage{graphicx}
\usepackage{textcomp}
\usepackage{xcolor}
\def\BibTeX{{\rm B\kern-.05em{\sc i\kern-.025em b}\kern-.08em
    T\kern-.1667em\lower.7ex\hbox{E}\kern-.125emX}}
\let\ssstyle=\scriptscriptstyle
\def\sT{{\ssstyle T}}
\usepackage{bm}
\graphicspath{ {figures/} }    
\renewcommand{\baselinestretch}{0.97}
\begin{document}

\title{28~GHz mmWave Channel Measurements: A Comparison of Horn and Phased Array Antennas and Coverage Enhancement Using Passive and Active Repeaters
\thanks{This work has been supported in part by NASA under the Federal Award ID number NNX17AJ94A and by DOCOMO Innovations, Inc.}
}
\author{\IEEEauthorblockN{Ozgur Ozdemir$^1$,~Fatih Erden$^1$,~Ismail Guvenc$^1$, Taha Yekan$^2$, and Tom Zarian$^2$}
\IEEEauthorblockN{$^1$\textit{Dept. Electrical \& Computer Eng., NC State University}, Raleigh, NC}
\IEEEauthorblockN{$^2$\textit{Metawave Corporation}, Palo Alto, CA} 
\IEEEauthorblockN{Email: \{oozdemi, ferden, iguvenc\}@ncsu.edu, \{taha.yekan, tom.zarian\}@metawave.co}}
\maketitle

\thispagestyle{plain}
\pagestyle{plain}

\begin{abstract}
Propagation channel characteristics are substantially different in sub-6~GHz and millimeter-wave (mmWave) bands. A typical mmWave link experiences more than an order-of-magnitude larger path loss and is more susceptible to blockages than a traditional sub-6~GHz link. Therefore, extensive measurement campaigns with efficient channel sounders are needed for a complete understanding of the mmWave propagation channel characteristics. In addition, solutions should be sought to overcome the outage problems associated with the mmWave bands and to achieve acceptable communication performance. In this paper, our 28~GHz channel sounder is introduced that can be used with horn antennas as well as phased array antennas. The two antenna types are then compared in terms of speed they switch from one direction to another and extracted multipath component~(MPC) properties, such as path gain. We then propose the use of passive reflectors and an active repeater for coverage enhancement by improving the received signal strength. In particular, ECHO reflector and TURBO repeater are introduced, and measurement results using them are presented.\looseness=-1

\end{abstract}

\begin{IEEEkeywords}
28~GHz, millimeter-wave (mmWave), phased array, reflector, repeater, rotated directional antenna. 
\end{IEEEkeywords}

\section{Introduction}

The rollout of a new 5G mmWave network to support the growing bandwidth and lower latency needs of the typical end-user is an enormous task. High data rates requirements in fifth generation~(5G) cellular systems can be met using large bandwidths on the order of GHz~\cite{Shafi2018}. As sub-6~GHz spectrum is currently overutilized, higher-frequency bands, such as millimeter-wave~(mmWave) bands, have attracted growing attention among the researchers. In order to deploy wireless systems, an accurate understanding of the channel propagation characteristics in the deployment band is required. Extensive channel measurement campaigns were performed to derive statistical channel models for sub-6~GHz cellular systems. However, as characteristics of the wireless channel are different at different frequencies, similar efforts should also be put in place to characterize the nature of radio propagation at mmWave frequencies~\cite{khatun&globalsip:2018,fatih_library}.

To address the problem of high path loss in mmWave systems, high gain directional antennas are used, and the antennas are aligned to improve the received signal strength. The channel sounders used for modeling the sub-6~GHz channel mainly measure the power delay profile~(PDP) of the channel. However, due to the high path loss in the mmWave bands, angular profile of the channel, known as the power angular-delay profile~(PADP), should also be measured~\cite{Lin2017} in these bands. \looseness=-1

We use horn antennas as well as phased array antennas to characterize the mmWave channel at 28~GHz. Although both of these antennas are directional, the horn antennas are mechanically steered whereas the amplitudes and phases of the phased array antennas are electrically switched to steer the beam from one direction to another. We compare the performance of the two antennas when used on our mmWave channel sounder in the exact same environment.

One of the most significant obstacles wireless providers face is extending 5G radio ranges and providing service into dead zones (e.g., inside the buildings, under bridges, behind structures) and “bend signals” around corners to connect to backhaul radios. Today carriers are forced to use expensive, delay adding repeater network elements that, in the digital domain, terminate and recreate the signals to solve these issues. Given these obstacles, innovative, cost-effective, low latency network reflectors and repeaters are critical. Deploying these platforms will greatly increase network capacity and overall coverage, and will also significantly lower operating and capital costs, giving enterprises and consumers reliable 5G coverage. Passive reflectors have been studied in the past for long distance satellite communications~\cite{NASA_refl,Literature4,Literature5} and for downlink communications~\cite{Microwave_refl,Literature6}. To increase coverage in mmWave systems, reflectors were considered in~\cite{lit_60GHz_indoor,Literature3,Literature1, indoor_wahab, wahab_outdoor}. 

In this paper, we also introduce ECHO passive reflector and TURBO active repeater from Metawave Inc, as solutions to coverage problems in mmWave communication. Metawave offers mmWave 5G platforms that enable higher speeds, provide latency, enable wider coverage, and offer non-line-of-sight~(NLOS) connectivity, using fewer 5G radios in cities, office buildings and malls, crowded stadiums and concerts, and other areas with 5G coverage. Measurements performed with ECHO and TURBO are presented. 

The remaining of the paper is organized as follows. In Section~\ref{Sec:Measurement}, we introduce the measurement hardware and give a high-level description of their operating principles. In Section~\ref{Sec:PADP}, we provide the PADP model, which we use to analyze the channel measurements. We compare the results obtained with the horn antennas and the phased array antennas in Section~\ref{Sec:RDAvsPhasedArr} and present the improvements in coverage obtained with the reflectors and the active repeater in Section~\ref{Sec:ReflectorResults} and Section~\ref{Sec:RepeaterResults}, respectively. Finally, we give concluding remarks in Section~\ref{Sec:Conclusion}.

\section{Measurement Hardware}
\label{Sec:Measurement}
In this section, we describe the mmWave channel sounder used in our measurements and introduce the ECHO reflector and TURBO repeater.

\subsection{mmWave Channel Sounder}
\label{Sec:RDA}
\begin{figure}[t!]
\centering
\centerline{\includegraphics[width=\linewidth]{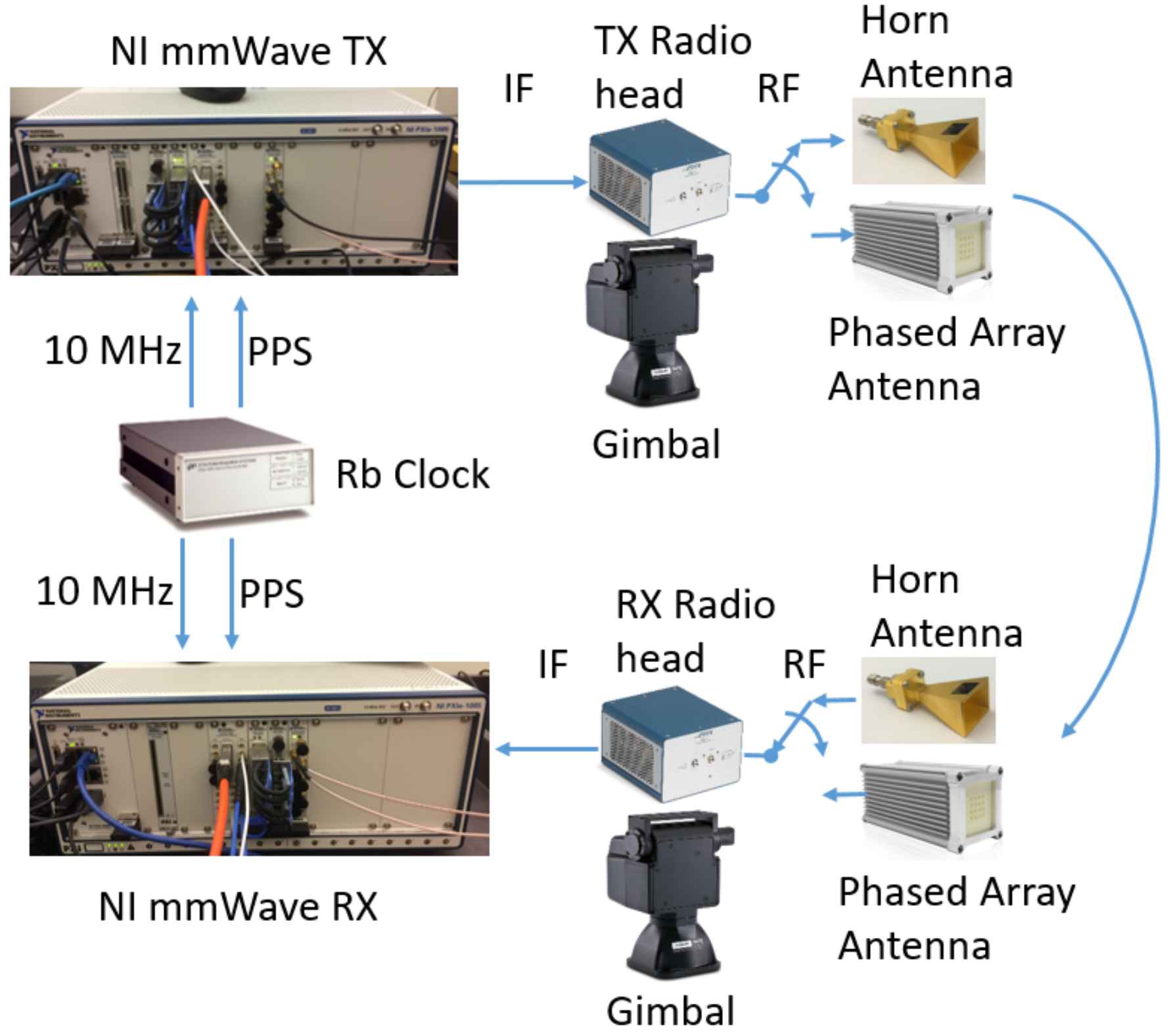}}
\caption{28~GHz mmWave channel sounder TX and RX hardware setup with single Rb clock. Either horn antennas or phased array antennas are connected to the mmWave radio heads.}\label{Fig:pxisetup}
\vspace{-3mm}
\end{figure}

The measurements are performed using an NI-based channel sounder~\cite{NImmwave} shown in Fig.~\ref{Fig:pxisetup}. The sounder hardware consists of NI PXIe-1085 TX/RX chassis, 28~GHz TX/RX mmWave radio heads from NI, FS725~Rubidium~(Rb) clocks~\cite{SRS}, and FLIR~PTU-D48E gimbals~\cite{FlirSystems}. The 10~MHz and pulse per second~(PPS) signals generated by a single rubidium~(Rb) clock are connected to PXIe~6674T timing modules at the TX and the RX. For measurements that the TX and RX need to be separated by more than 30~m, two separate Rb clocks are used at the TX and the RX. In that case, one of the clocks trains the other clock for at least one hour before the measurements so that the clocks are synchronized. 

The LabVIEW-based sounder code periodically transmits a Zadoff-Chu (ZC) sequence of length 2048 to sound the channel. The ZC sequence is over-sampled by 2 when the sounder operates on 2~GHz bandwidth. Then, it is filtered by the root-raised-cosine filter, and the generated samples are uploaded to PXIe-7902 FPGA. These samples are sent to PXIe-3610 digital-to-analog converter with a sampling rate of $f_s=3.072$~GS/s. The PXIe-3620 module up-converts the base-band signal to IF at 10.56~GHz, and the mmWave radio head up-converts the IF signal to 28~GHz RF. This process is reversed at the RX side, and the complex CIR samples are sent to the PXIe-8880 host PC for further processing. The channel sounder provides 0.651~ns resolution in the delay domain. The dynamic range of the analog-to-digital converter is $60$~dB, and the maximum measurable path loss is $185$~dB.

The mmWave channel sounders typically measure the angular profile of the channel in addition to the power delay profile to obtain the PADP. In order to measure the PADP of the channel, directional antennas are used. In our channel sounder, we use either a horn antenna or a phased array antenna as shown in Fig.~\ref{Fig:pxisetup}.

The directional horn antennas we use~\cite{sageM} have 17~dBi gain and $26^{\circ}$/$24^{\circ}$ beam-width in the elevation/azimuth plane. The radio heads are placed on the rotating gimbals to be able to measure the PADP of the channel. The gimbals rotate the antennas in every possible direction to scan the entire azimuth and elevation planes and to perform a separate measurement in each direction. The total time required for the entire measurement depends on the number of positions visited by the antennas as well as the speed of the gimbal that rotates the horn antenna mechanically. The mechanical rotation is usually slower than electrical rotation as in phased array antennas. When fast measurements are needed, phased array antennas are preferred although they are more expensive, more difficult to calibrate, and they have limited field of view. The phased array antenna used in our we use~\cite{bbox} have 39~dBi transmitter gain and 21.5~dBi receiver gain. It operates in 24 to 31 GHz frequency range. The direction of this phased array antenna can be switched in 15 ms. This gives us a significant improvement over the mechanical switching of the horn antennas in terms of total measurement time. A more detailed comparison of horn antenna and phased array antenna, when they both measure the same environment, will be provided in Section~\ref{Sec:RDAvsPhasedArr}. 

\begin{figure*}[!t]
\centering
\subfloat{\raisebox{12mm}{\includegraphics[trim=7.5cm 5cm 8.5cm 5cm,height=3.3cm,width=0.4\linewidth]{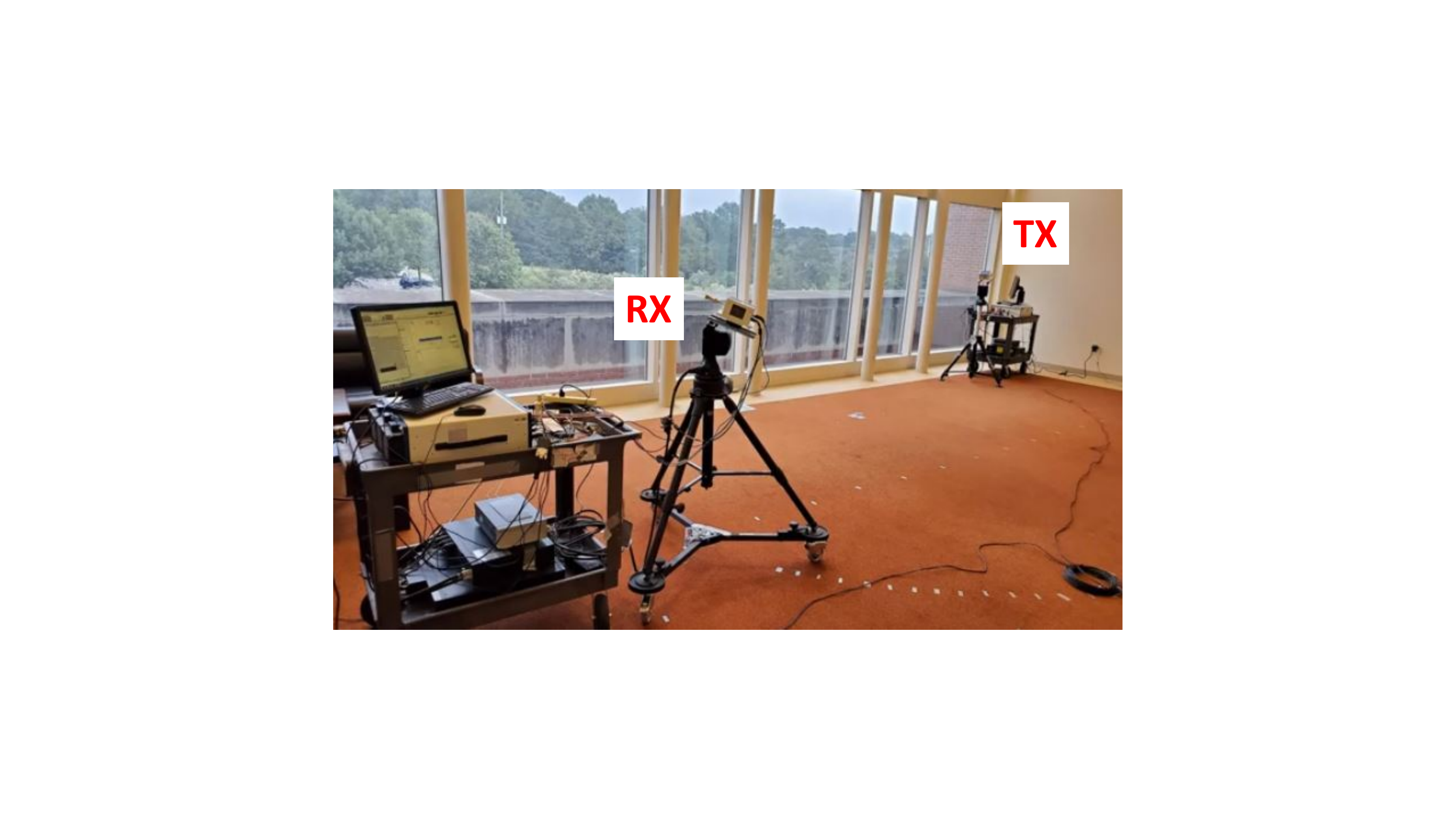}}
\label{Fig:photo_RDA_vs_phasedArr)}}
\hspace{10pt}
\subfloat{\includegraphics[trim=2cm 1.5cm 16cm 1.1cm, clip, width=0.31\linewidth]{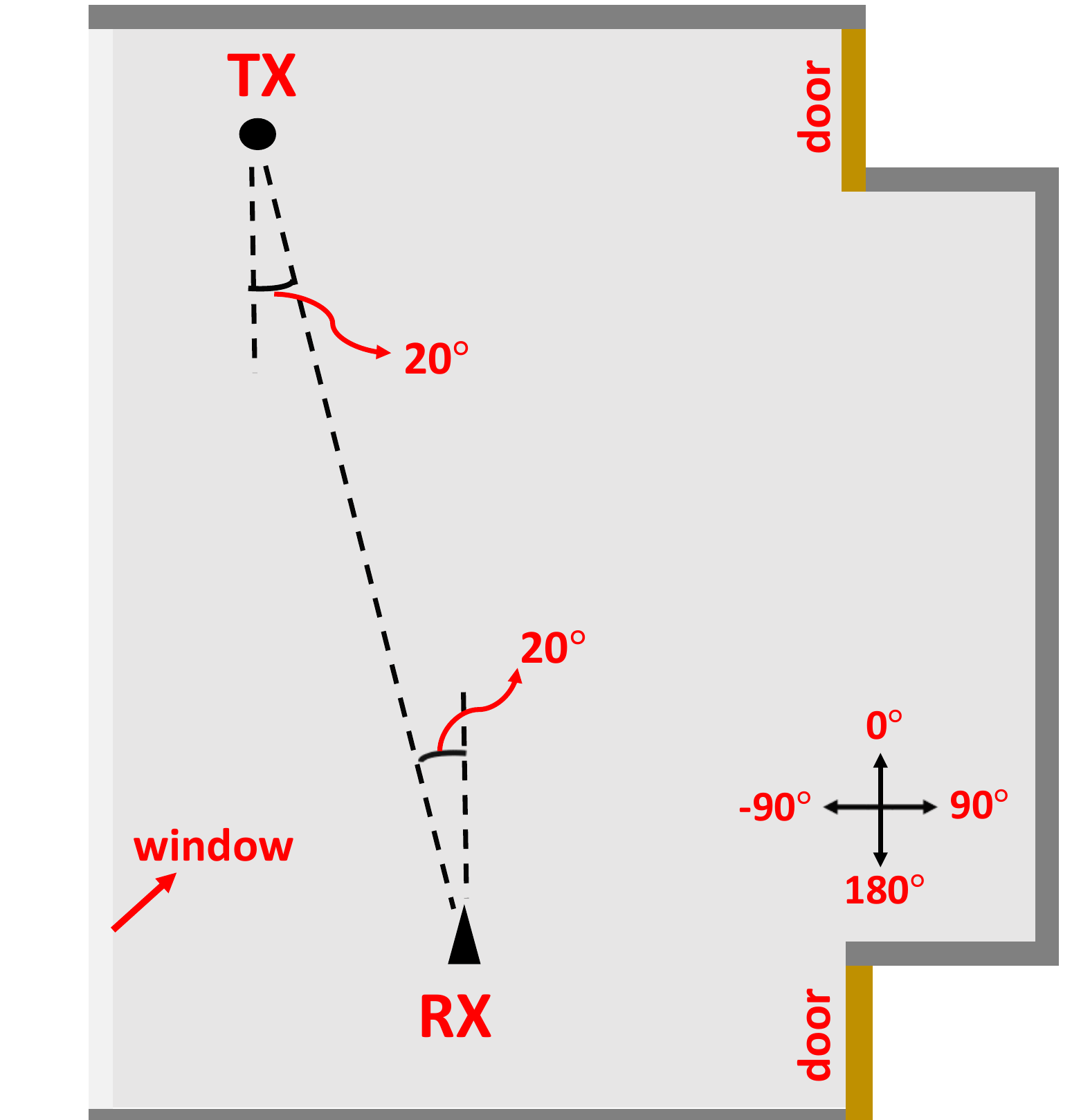}
\label{Fig:sketch_RDAvsphasedArr}}
\caption{Measurement setup in a seminar hall at NCSU campus for comparing measurements with horn antennas and phased array antennas.}\label{Fig:setup_RDA_vs_phasedArr}
\vspace{-3mm}
\end{figure*}

\subsection{ECHO Reflector and TURBO Repeater from Metawave}
\label{Sec:reflector}
One possible solution to improve coverage by increasing received signal strength in mmWave systems is using metallic reflectors. In an earlier study~\cite{wahab_outdoor}, the use of passive reflectors in an outdoor setting was proven to be helpful to extend the coverage to areas that are blocked by obstacles. In this study, we will look into a similar but more effective solution, the ECHO reflector. ECHO is a Metawave solution to illuminate dead zones for 5G FR2 bands. It acts as a passive beamformer to capture the power from base station and spread it to desired spaces that 5G signal cannot reach. It improves the network performance and extends the coverage. It is designed to support both H/V polarizations and can be designed wideband to cover wider frequency range. The antenna is designed based on information like location and specifications of base station, user equipment, interested frequency band, and etc. The product is planar and conformal so that it can easily be installed on walls and other places without needing much maintenance and no requiring to get license for its operation as it does not need any electricity to power up. The antenna is fabricated based on standard process by using common laminates, making the production price low.  


Metawave provides carriers with the world’s first all-analog, low-cost, active repeaters for faster deployment and effective indoor and outdoor 5G coverage. Metawave TURBO Active Repeaters boost signal strength over a thousand times to greatly extend coverage in challenging areas. 
Metawave is designing its second-generation TURBO Active Repeater, able to be remotely configured with little installation effort by carriers. TURBO will be equipped with remote access capabilities for monitoring and reporting the health of the unit. Equipped with dual polarized reconfigurable antenna arrays, TURBO will deliver even faster connectivity than our current platform, which is being tested by carriers around the world.

\section{Channel PADP Modeling}
\label{Sec:PADP}
Our mmWave channel sounder  measure the PADP of the wireless channel from CIR measuremnts at a given position of the transmitter and the receiver. The PADP of the channel to be measured can be expressed as: 
\begin{align}
\label{PADP:eq}
PADP(\tau, \bm{\theta}^{\mathrm{AoD}} , \bm{\theta}^{\mathrm{AoA}}) = &\sum_{n=1}^{N}  \alpha_n \delta (\bm{\theta}^{\mathrm{AoD}}-\bm{\theta}^{\mathrm{AoD}}_n)  \\ \nonumber &\times  \delta (\bm{\theta}^{\mathrm{AoA}}-\bm{\theta}^{\mathrm{AoA}}_n) \delta(\tau-\tau_{n}),
\end{align}
where $\bm{\theta}^{\mathrm{AoD}}_n = [\theta^{\mathrm{AoD,Az}}_n \,\, \theta^{\mathrm{AoD,El}}_n]^{\sT}$ is the two-dimensional AoD of the $n$th MPC at the TX in the azimuth and elevation planes, $\bm{\theta}^{\mathrm{AoA}}_n = [\theta^{\mathrm{AoA,Az}}_n \,\, \theta^{\mathrm{AoA,El}}_n]^{\sT}$ is the two-dimensional AoA of the same MPC at the RX, $\alpha_n$ is the path gain, $\tau_n$ is the delay of the $n$th MPC, and $N$ is the total number of MPCs considering all possible TX/RX directions.

At the $m$th position of the TX/RX antennas $\bm{\theta}^{\mathrm{TX}}_m = [\theta^{\mathrm{TX,Az}}_m \,\, \theta^{\mathrm{TX,El}}_m ]^{\sT}$ and  $\bm{\theta}^{\mathrm{RX}}_m = [\theta^{\mathrm{RX,Az}}_m \,\, \theta^{\mathrm{RX,El}}_m ]^{\sT}$ denote the angles of the TX and the RX antennas, respectively. At this position the channel sounder measures the CIR expressed by
\begin{align}
\label{CIR:eq}
h_m (\tau) = \sum_{n=1}^{N} & \sqrt{\alpha_n  G_{\mathrm{TX}}(\bm{\theta}^{\mathrm{TX}}_m-\bm{\theta}^{\mathrm{AoD}}_n) G_{\mathrm{RX}}(\bm{\theta}^{\mathrm{RX}}_m-\bm{\theta}^{\mathrm{AoA}}_n)} \\ \nonumber &\times  \mathrm{e}^{j \phi_n} \delta(\tau-\tau_n)+w_m(\tau),
\end{align}
where the TX/RX antenna gains, which are functions of the antenna angles, are also considered as part of the CIR.
Here, $G_{\mathrm{TX}}(\cdot)$ and $G_{\mathrm{RX}}(\cdot)$ are the TX and the RX antenna gains, $\phi_n$ is the phase of the $n$th MPC, and $w_m(\tau)$ is the noise in the $m$th measurement. In~\cite{erden2019}, we explain in detail the process of obtaining PADP parameters $\alpha_n$, $\tau_n$, $\bm{\theta}^{\mathrm{AoD}}_n$, and $\bm{\theta}^{\mathrm{AoA}}_n$ as introduced in (\ref{PADP:eq}) from CIR measurements $h_1 (\tau)$ ... $h_M (\tau)$, where $M$ is the total number of angles scanned during measurements. \looseness=-1



\section{PADP Measurements: Horn Antenna vs Phased Array Antenna}
\label{Sec:RDAvsPhasedArr}

To compare the horn antenna vs phased array antenna PADP measurements were performed in a seminar hall at NCSU campus as shown in Fig.~\ref{Fig:setup_RDA_vs_phasedArr}. During the measurements 3 different angles $\{-20^\circ, 0^\circ, 20^\circ\}$ were scanned by the transmitter and the receiver antennas in the elevation planes. In the azimuth plane 17 different angles from the set $\{-180^\circ, -160^\circ, \dots, 0^\circ, \dots, 160^\circ\}$ were scanned skipping $0^\circ$ for the transmitter and $-180^\circ$ for the receiver as the gimbal can not rotate whole $360^\circ$ in the azimuth plane. The total number of channel measurements $M$ is, therefore, $3 \times 17 \times 3 \times 17=2601$. The phased array antenna is able scan three elevation angles $\{-20^\circ, 0^\circ, 20^\circ\}$. In order to make a fair comparison, in the azimuth plane, the phased array antenna was rotated to the angles $\{-160^\circ, -100^\circ, 0^\circ, 100^\circ, 160^\circ \}$ on top of the gimbal. Then phased array antennas were able to scan the five azimuth angles in the set $\{-40^\circ, -20^\circ, \dots, 40^\circ\}$ relative to the position of the gimbal in addition to the elevation angles mentioned above. The $M$ measurements out of the total measurements were then used to extract PADP parameters as in the measurements with the horn antennas. 

The measurements started with the horn antennas. At each position out of the M positions, the CIR given in~(\ref{CIR:eq}) was measured and saved for post processing. Once the measurements with the horn antennas are completed the measurements with the phased array antenna were performed in the same environment. When performing the measurements with phased array antennas for fixed positions of the gimbals, the total number of channel measurements were $3 \times 5 \times 3 \times 5=225$. This is repeated for five different positions of the gimbals at the transmitter and at the receiver, therefore, the total antenna switching for the phased array antennas was $225 \times 25=5625$. 
To compare the total measurement time for the horn antennas and phased array antenna, we assume 1~s per measurement for the horn antennas and 15~ms per measurement for the phased array antennas. In this case measurements with horn antennas take around $2601 \times 1~\mathrm{s} =43$~min whereas the measurements with phased array antennas take $5625 \times 15~\mathrm{ms} =1.4$~min. 


\begin{figure}[!t]
\centering
\subfloat[]{\includegraphics[width=0.80\linewidth]{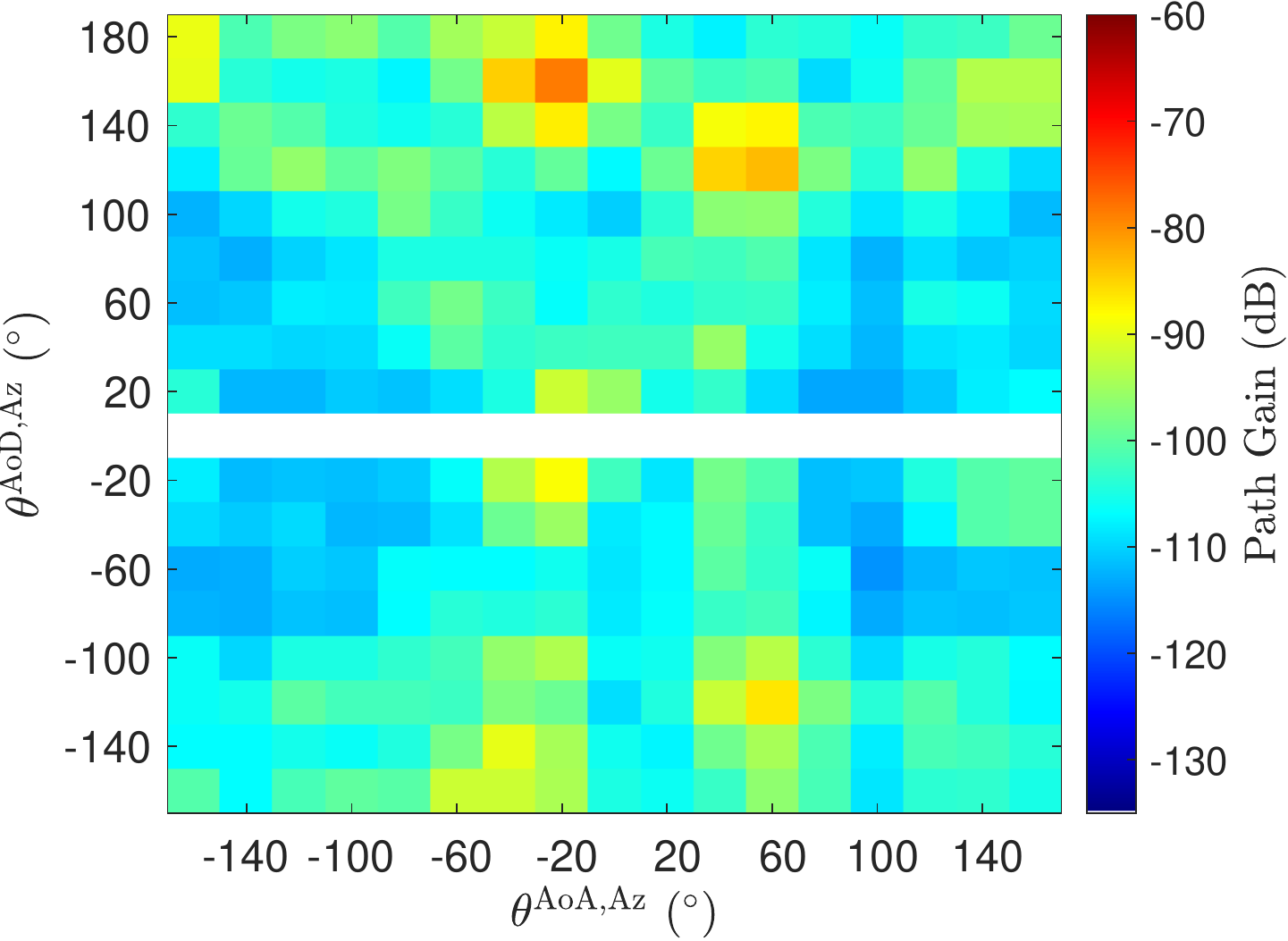}
\label{fig:PG_RDA}}
\hspace{0pt}
\subfloat[]{\includegraphics[width=0.80\linewidth]{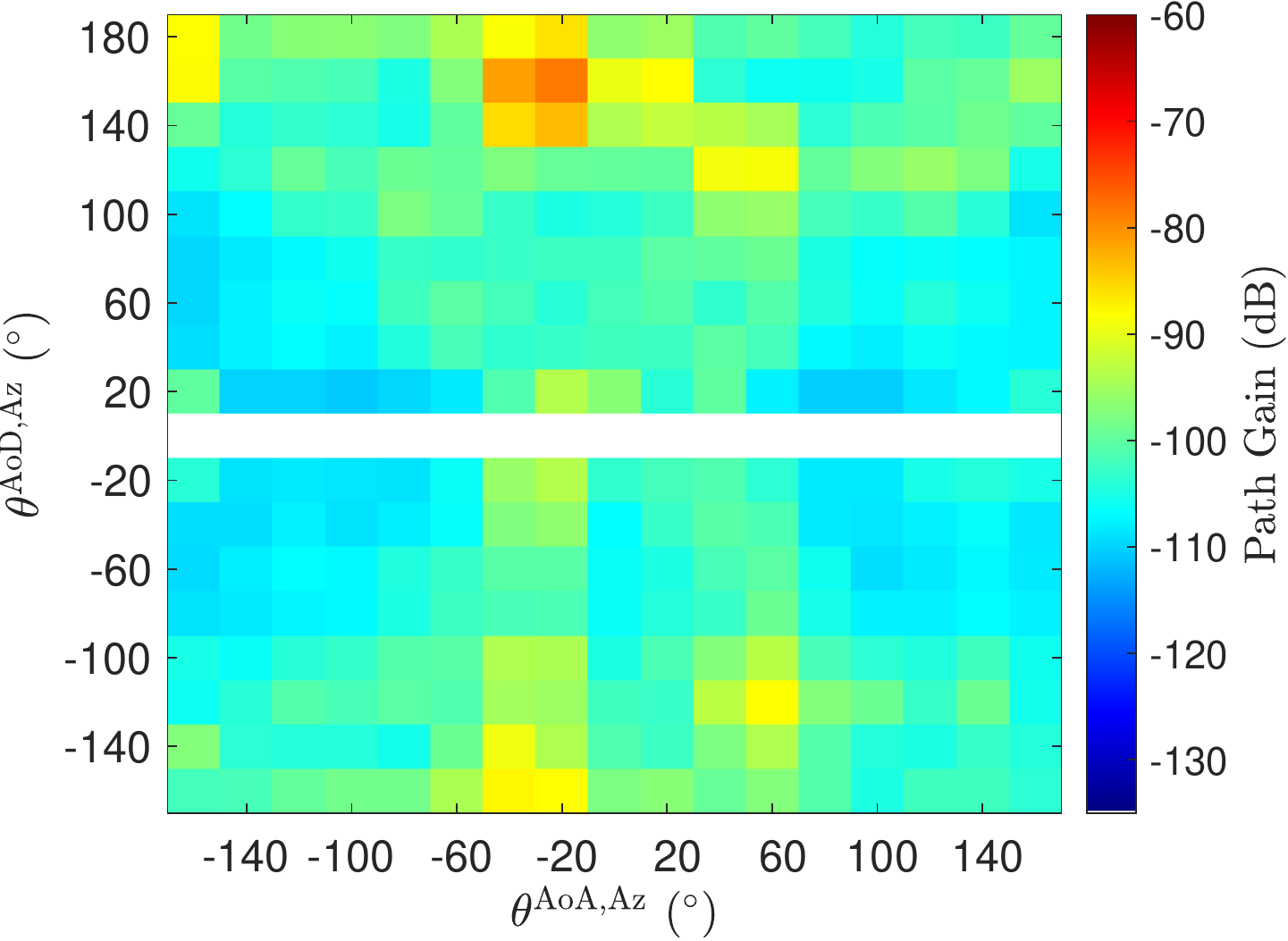}
\label{fig:PG_phasedArr}}
\caption{Path gain: (a) horn antenna and (b) phased array antenna. TX $El = 0^\circ$, RX $El = 0^\circ$.}
\label{fig:PG_El00}
\vspace{-3mm}
\end{figure}

By squaring the CIR measured at a given position, the PDP for that position can be obtained. Summing up all the powers of the PDP gives the total received power at that position. The total received power at a position can be used to calculate the path gain at that position as follows
\begin{equation}
\hat{\alpha}(m)=P_{\mathrm{RX}}(m)-P_{\mathrm{TX}}-G_{\mathrm{TX}}-G_{\mathrm{RX}},    
\end{equation}
where $P_{\mathrm{TX}}$ is the transmit power, $P_{\mathrm{RX}}$ is the received power, and $G_{\mathrm{TX/RX}}$ is the TX/RX antenna gain. Fig.~\ref{fig:PG_El00} is a comparison of horn antenna (Fig.~\ref{fig:PG_El00}(a)) and phased array antenna (Fig.~\ref{fig:PG_El00}(b)) in terms of total path gain as a function of transmitter and receiver antenna positions in the azimuth plane keeping the elevation angles to be zero. Note that, path gain is a negative number which corresponds to a positive path loss. A close match between the two measurements corresponding to horn antenna and phased array antenna is observed. As there are 17 angles scanned in the azimuth plane, each of the plots contain $17 \times 17 \times=289$ measurements.

\begin{figure}[!t]
\centering
\subfloat[]{\includegraphics[width=0.75\linewidth]{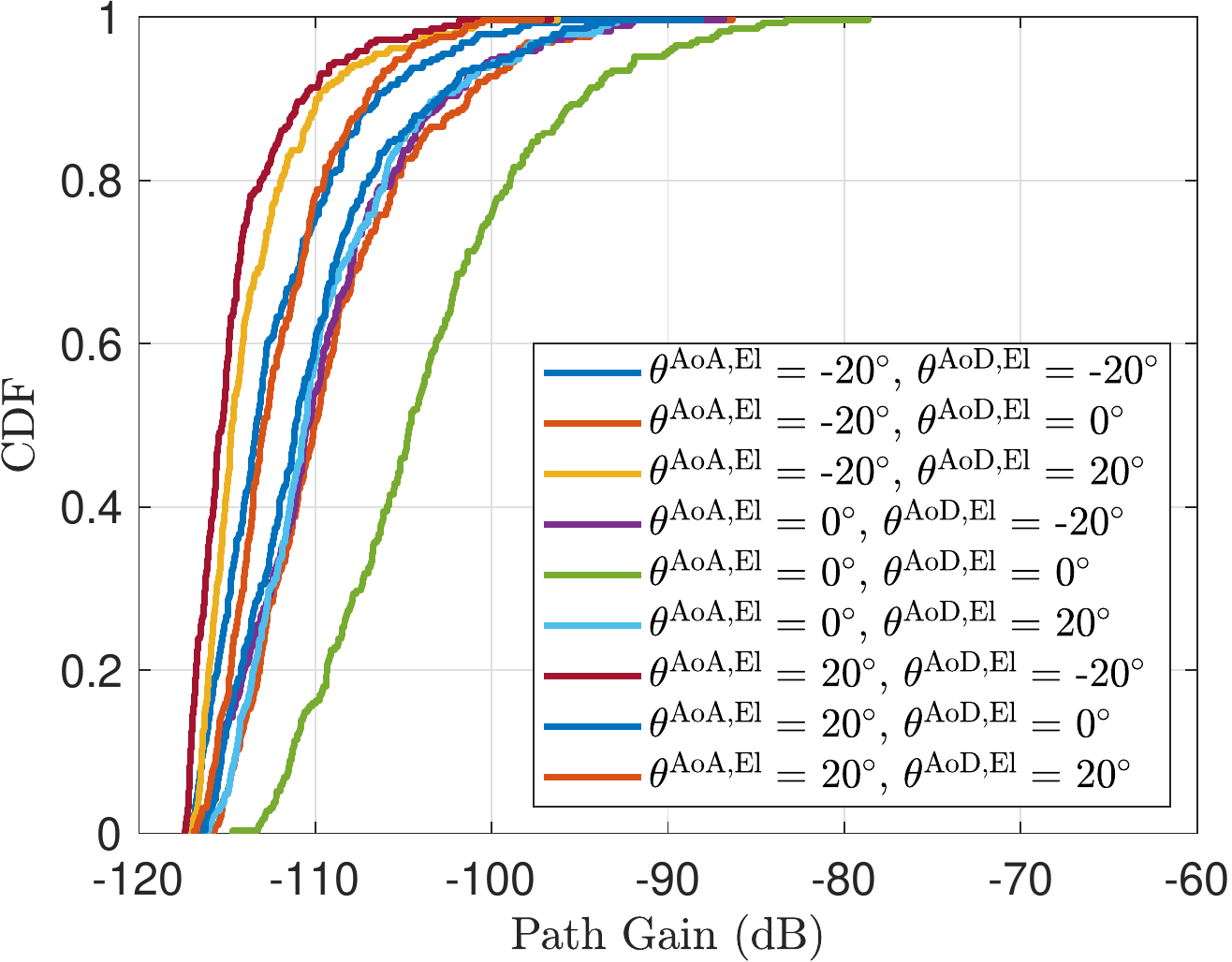}
\label{fig:CDF_RDA}}
\hspace{0pt}
\subfloat[]{\includegraphics[width=0.75\linewidth]{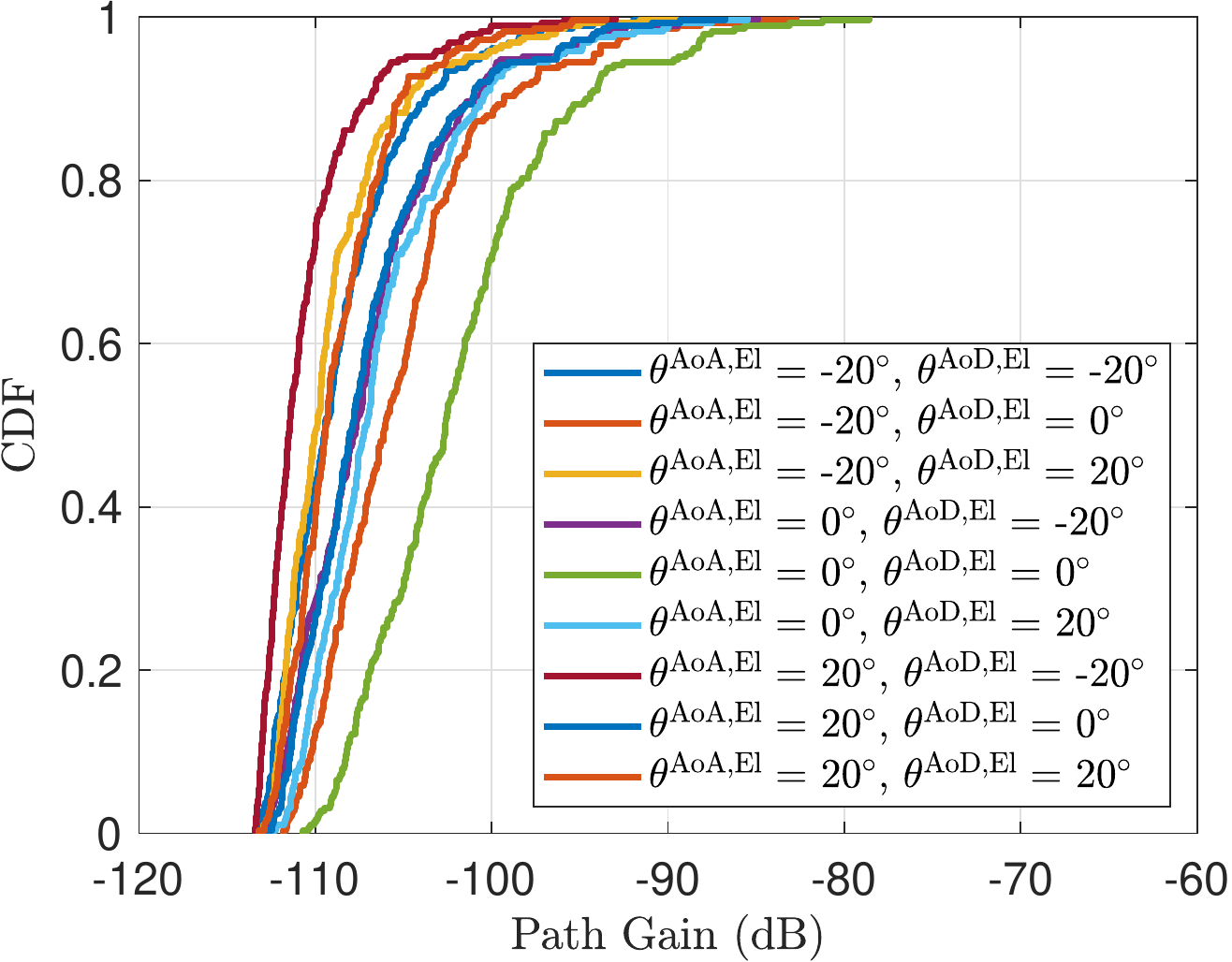}
\label{fig:CDF_phasedArr}}
\caption{CDF of the path gain: (a) RDA and (b) phased array.}
\label{fig:CDFs_pathGain}
\vspace{-3mm}
\end{figure}

The cumulative distribution function~(CDF) of the path gains at different azimuth angles are plotted in Fig.~\ref{fig:CDFs_pathGain}. Each plot corresponds to a fixed elevation angle at the transmitter and the receiver. The CDFs for horn antenna measurements are shown in Fig.~\ref{fig:CDFs_pathGain}(a) and the CDFs for phased array antenna measurements are shown in Fig.~\ref{fig:CDFs_pathGain}(b). Each plot contains nine CDF plots corresponding to combination of three different elevation angles at the transmitter and at the receiver. As expected when both elevation angles are zero we observe the highest path gains. We also note the close match between horn antenna and phased array antenna measurements.

\begin{figure}[!t]
\centering
\subfloat[]{\includegraphics[width=0.8\linewidth]{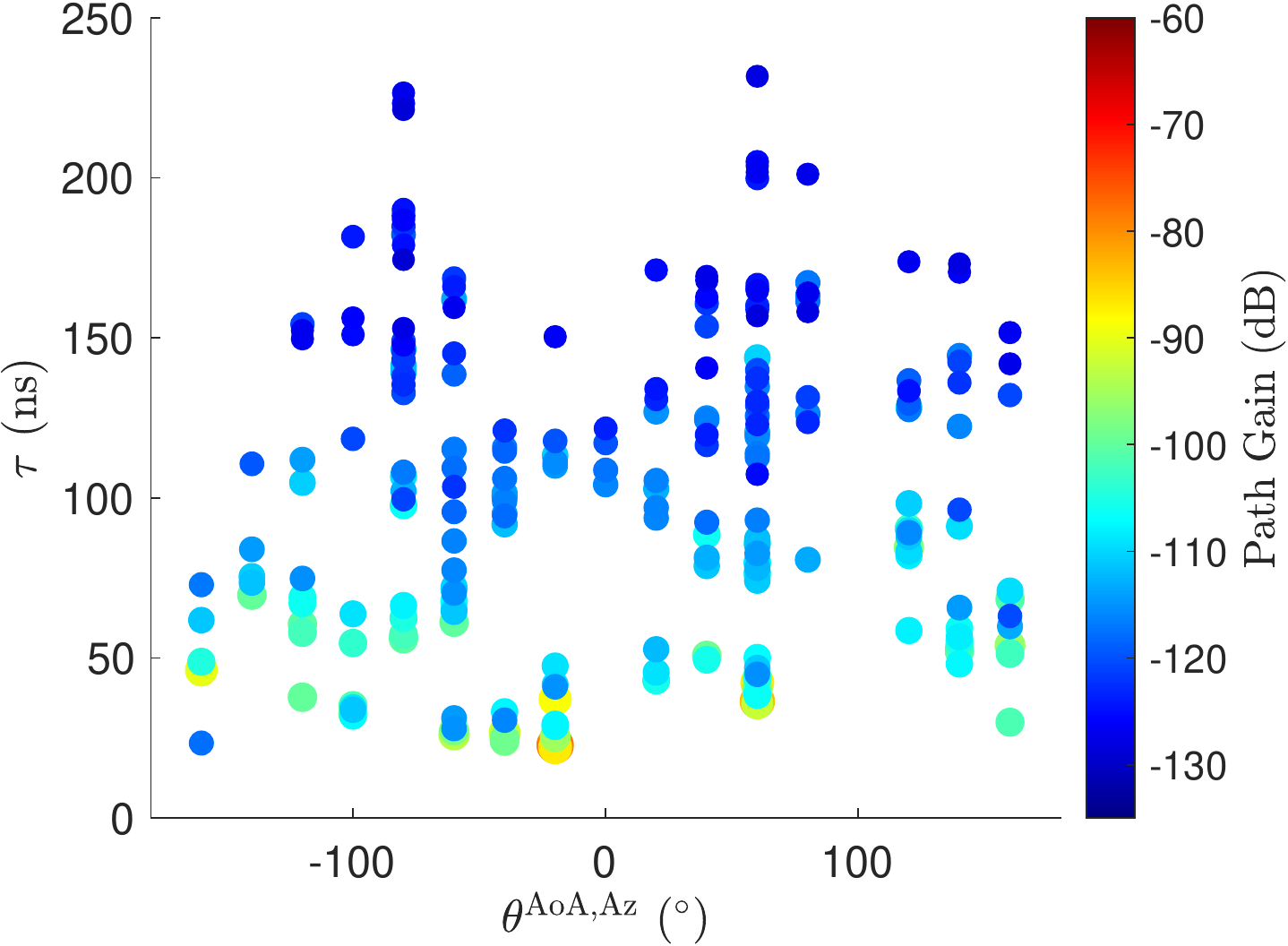}
\label{fig:MPC_RDA}}
\hspace{0pt}
\subfloat[]{\includegraphics[width=0.8\linewidth]{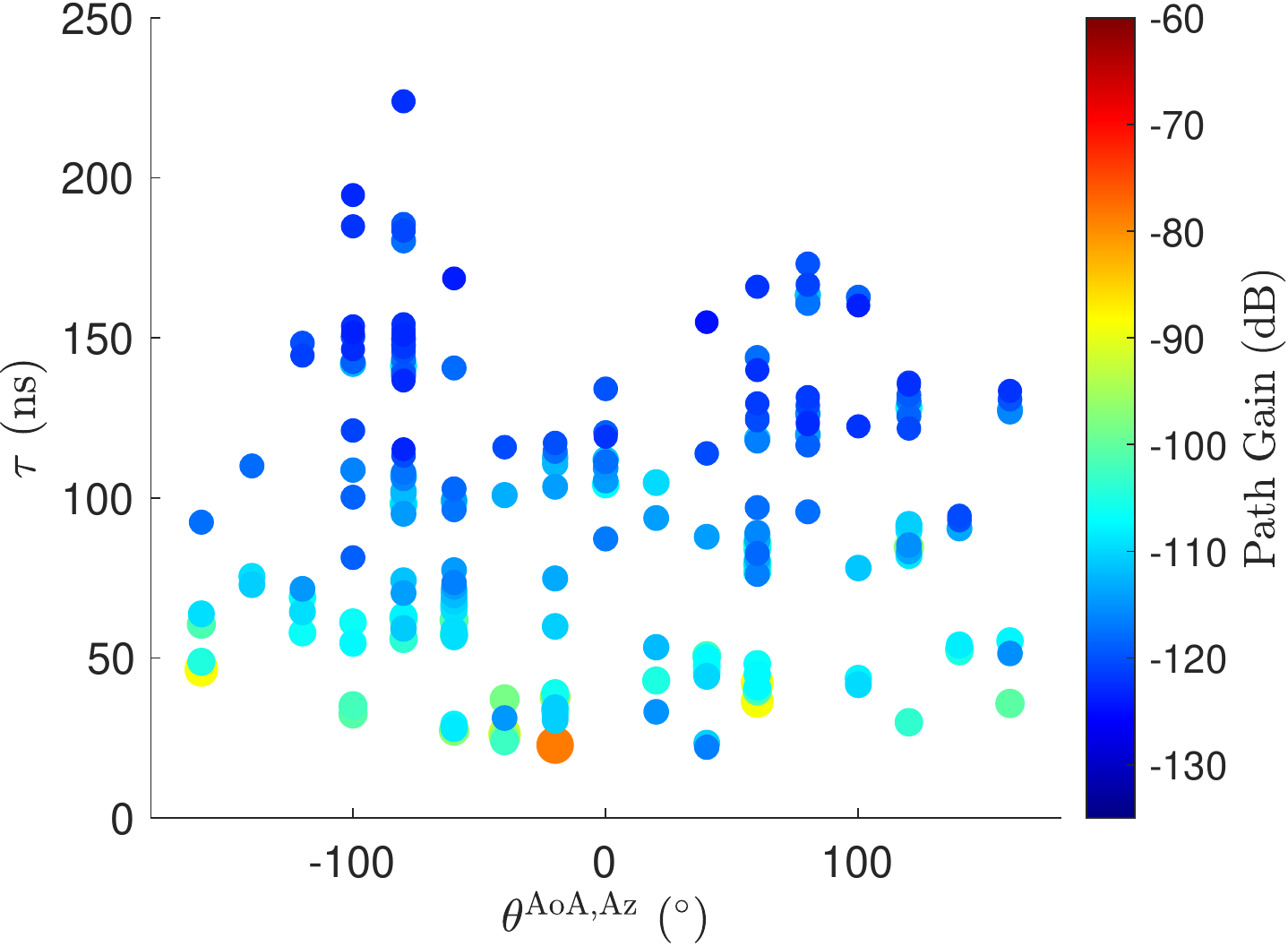}
\label{fig:MPC_phasedArr}}
\caption{MPCs extracted: (a) horn antenna and (b) phased array antenna.}
\label{fig:MPCs}
\vspace{-3mm}
\end{figure}

Once the measurements are performed and CIRs corresponding to each position of antennas are saved, we extract the MPCs using the method in~\cite{erden2019}. The extracted parameters of MPCs are path gain ($\alpha_n$), delay ($\tau_n$), AoD ($\bm{\theta}^{\mathrm{AoD}}_n$), and AoA ($\bm{\theta}^{\mathrm{AoA}}_n$) for each MPC $n$. Fig.~\ref{fig:MPCs} shows the AoA in azimuth plane, delay and path gains of all of the extracted MPC for horn antenna and phased array antenna measurements side by side for comparison. 

\section{Coverage Enhancement Using Metallic and ECHO Metawave Reflector}
\label{Sec:ReflectorResults}

Due to high path loss at mmWave frequencies, providing coverage especially for NLOS scenarios becomes difficult. Passive metallic reflectors can act as a repeater without electricity and minimal to no maintenance. Metallic reflectors reflect the signal in the same direction as the angle of incidence. On the other hand ECHO Metawave reflector is designed to reflect the incoming signal coming from a particular direction to a designed direction. This may be useful for situations where coverage at a particular direction is desired and reflector can only be placed in a particular orientation. In this section, we will present measurement results with an aluminum reflector and ECHO metawave reflector. The ECHO reflector we used in our measurements are designed to reflect signals coming at $52^\circ$ incident angle to $30^\circ$ reflection angle. 

\begin{figure}[t!]
\centering
\centerline{\includegraphics[width=0.7\linewidth]{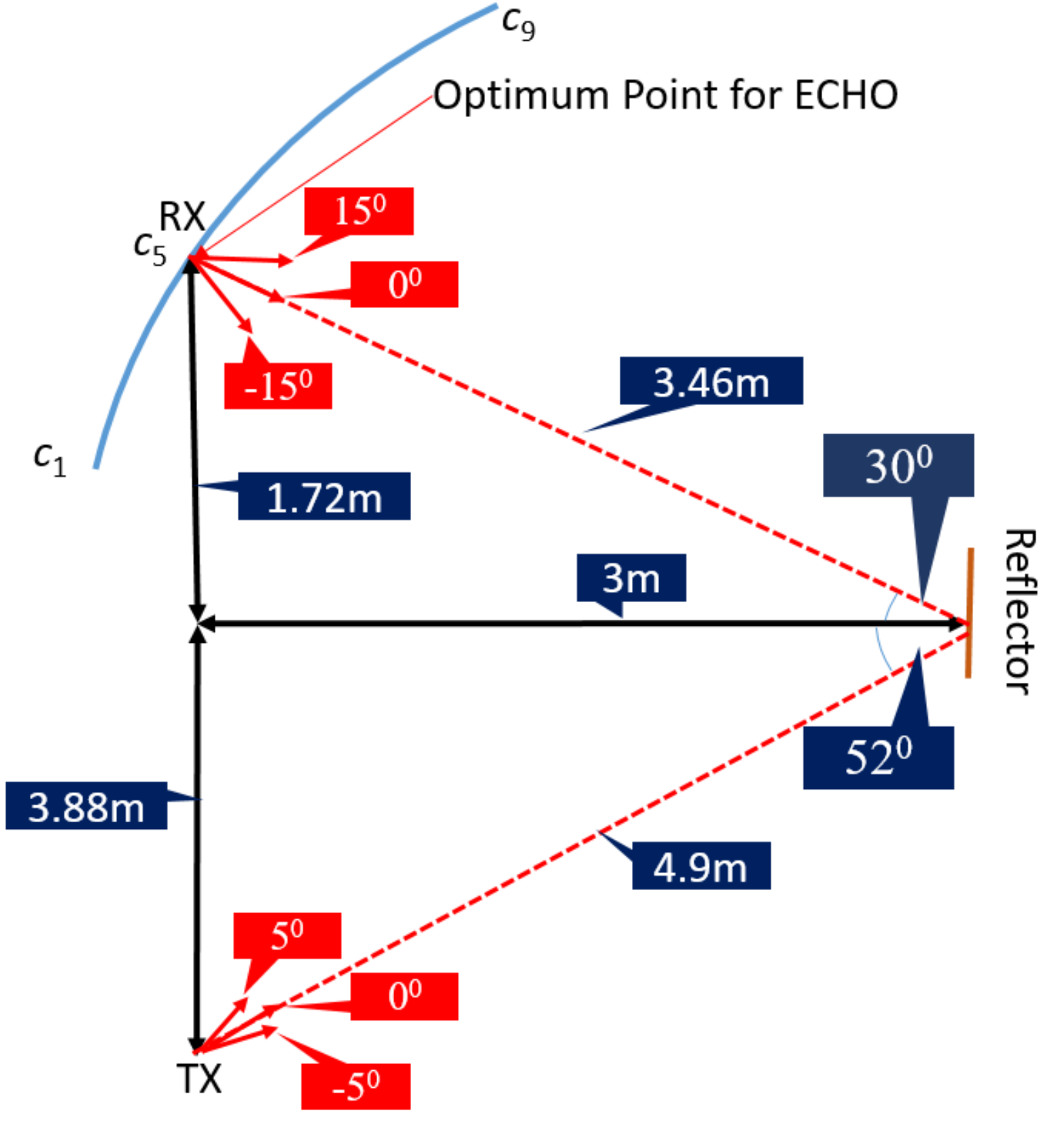}}
\caption{Seminar hall setup for ECHO measurements. The measurements are performed along a circle between points $c_1$ and $c_9$. Measurement points are separated by $8^{\circ}$.}
\label{fig:SHsetup}
\vspace{-3mm}
\end{figure}

\begin{figure*}[!t]
\centering
\subfloat[]{\includegraphics[width=0.33\linewidth]{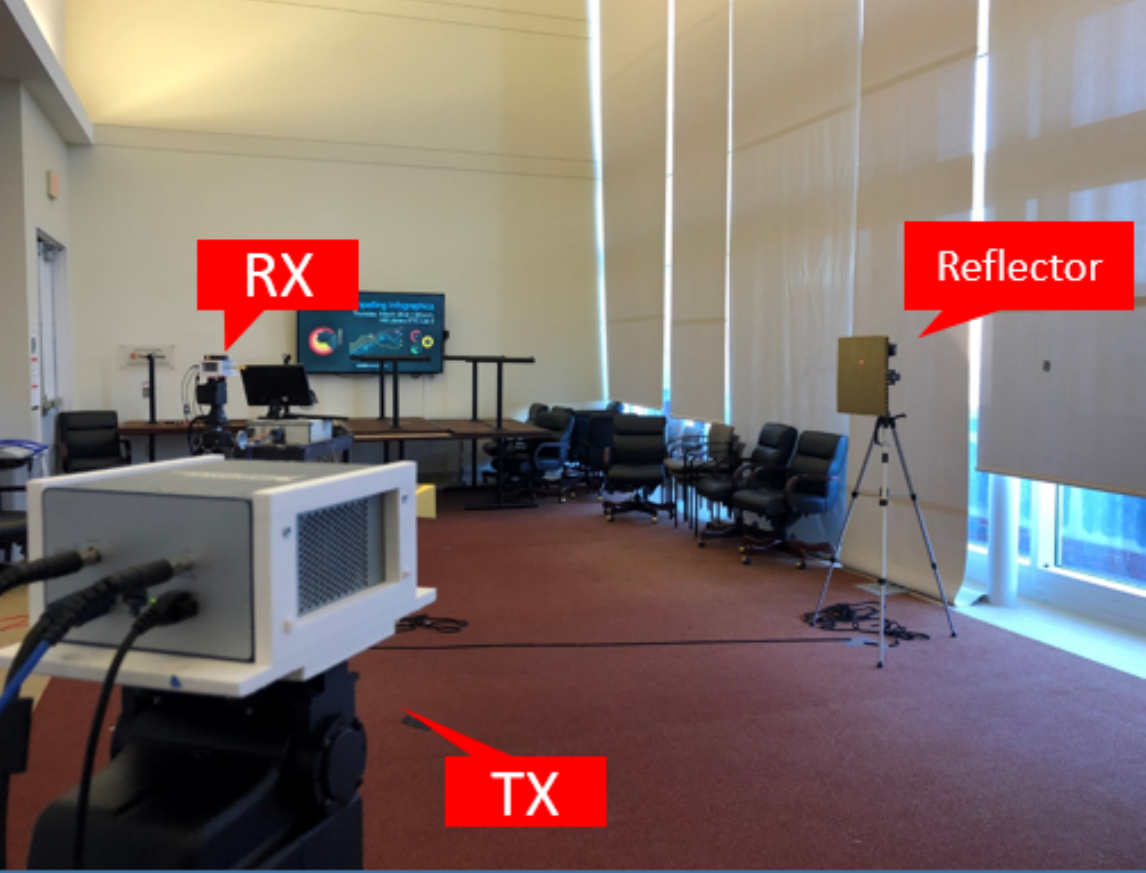}
\label{fig:indoorpicture}}
\hspace{13pt}
\subfloat[]{\includegraphics[width=0.33\linewidth]{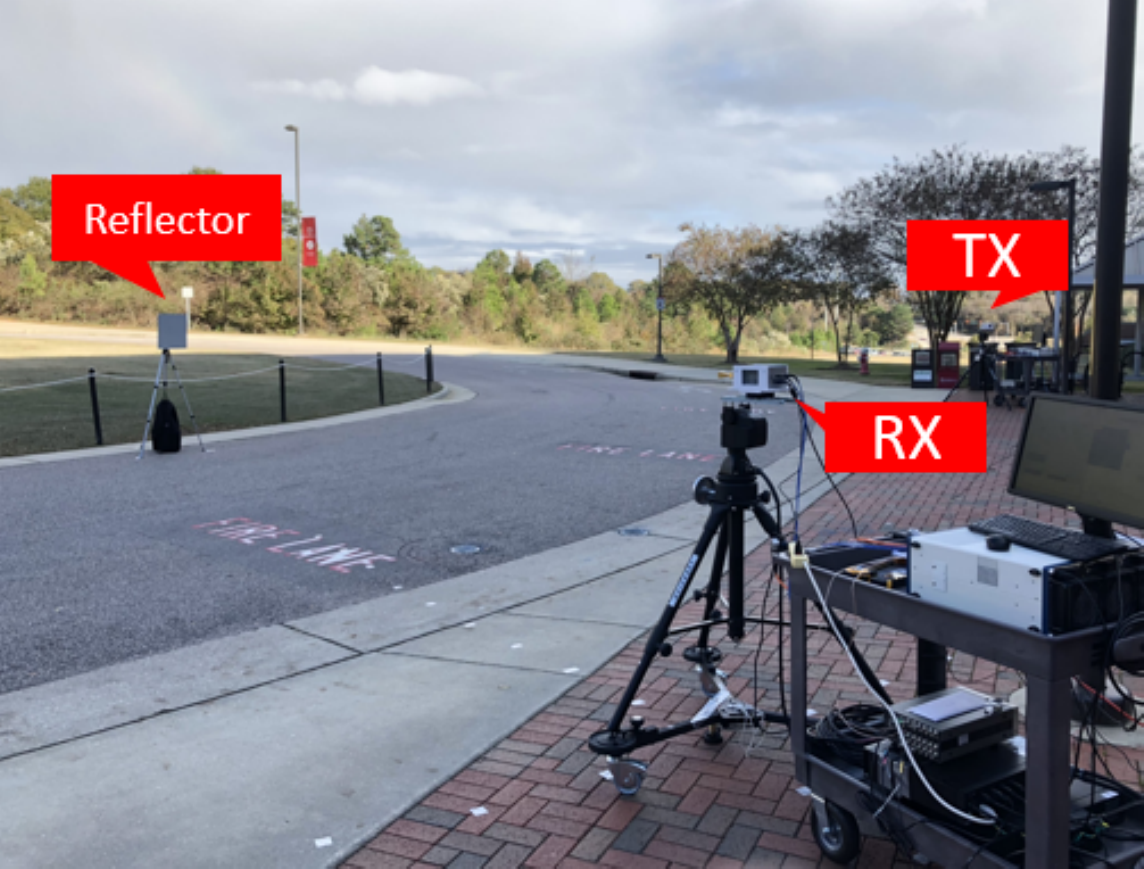}
\label{fig:outdoorpicture}}
\caption{(a) Indoor and (b) outdoor measurement environments for reflectors.}
\label{fig:pictures}
\vspace{-3mm}
\end{figure*}

The measurements were conducted in the seminar hall at NCSU campus and later repeated in outdoor as shown in Fig.~\ref{fig:pictures}. During the measurements the transmitter and the reflector is kept at a fixed position and the receiver is moved in a circle as illustrated in Fig.~\ref{fig:SHsetup}. The height of the transmitter, reflector, and the receiver are kept identical. The measurement positions for the receiver are selected so that when the receiver is at the point labelled as $c_5$, the angle of incident of the incoming ray on the reflector is $52^\circ$ and the signal towards the receiver reflects at $30^\circ$. Therefore, $c_5$ is the designed optimum point for the ECHO Metawave reflector. Performing measurements along a circle guarantees that the path from transmitter to the receiver reflecting off the reflector has the same distance and therefore the path loss is the same for all measurements. Prior to the measurements, the direction of the transmitter antenna is aligned to the reflector. For each position of the receiver, the direction of the receiver is also aligned to the reflector. The transmitter antenna scans the angles in the set $\{-5^\circ, -4^\circ, \dots, 5^\circ\}$ with $1^\circ$ increments. The receiver antenna scans the angles in the set $\{-15^\circ, -14^\circ, \dots, 15^\circ\}$ with $1^\circ$ increments. This is performed to correct for possible misalignment problems. \looseness=-1

At each position of the transmitter and the receiver CIR is measured and saved for post-processing. The total number of measurements at a given position is $11 \times 31 =341$. As the gimbal moves $1^\circ$ per measurement, it was possible to perform 1 measurement every 100~ms. As the measurement at a particular position takes around 0.5 min, horn antennas were used during measurements.

The measurement points $c_1$, $c_2$, $\dots$, $c_9$ are separated by $ 8^\circ$ i.e., the angle of reflection from reflector between the measurements points is $8^\circ$. The measurements were first performed with ECHO Metawave reflector and then repeated with identical sized aluminum reflector for comparison. For the ECHO Metawave reflector, three additional measurements between some of the two consecutive points were performed compared to the aluminum reflector to have better resolution.

\begin{figure}[t!]
\centering
\centerline{\includegraphics[width=\linewidth]{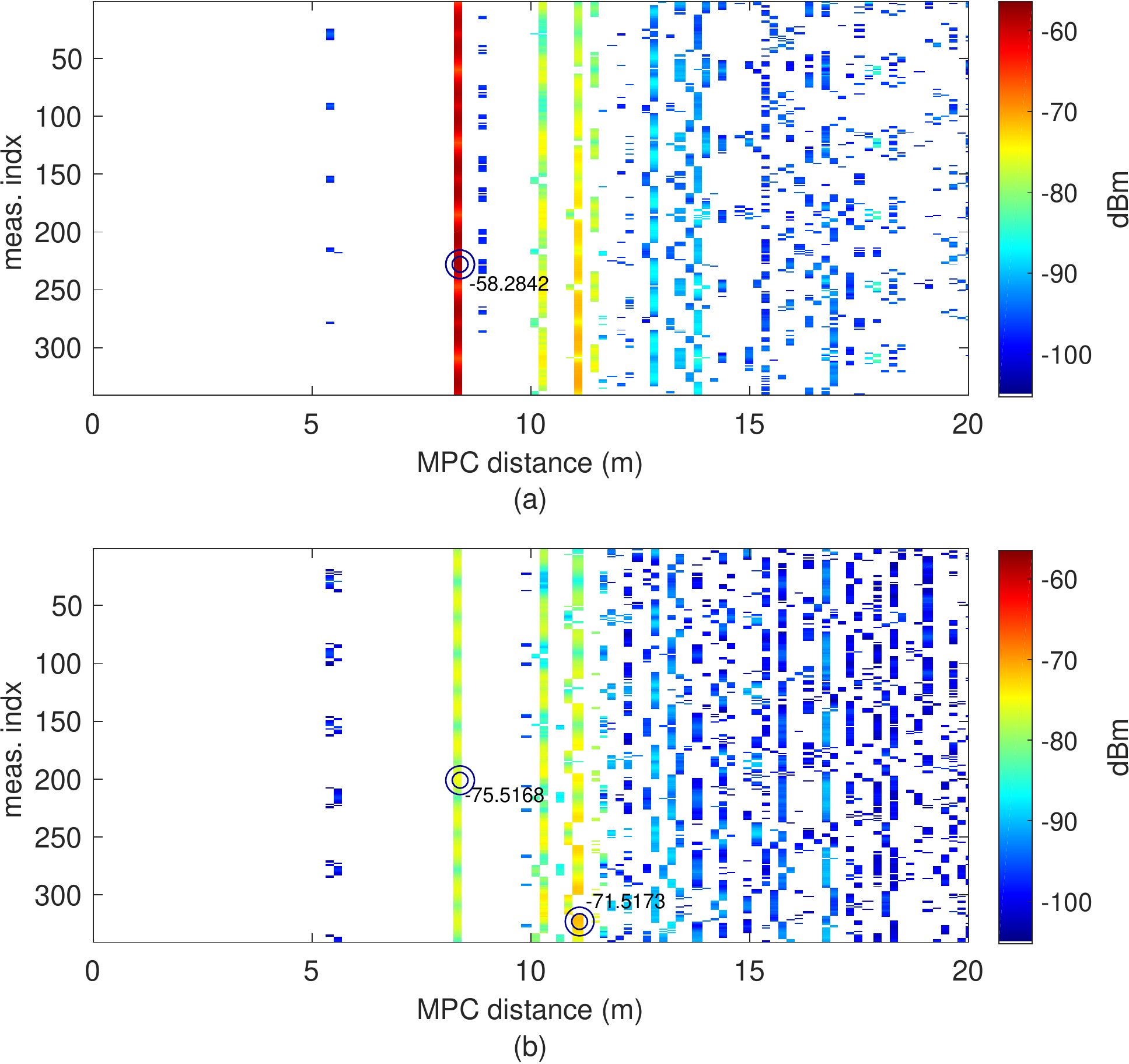}}
\caption{Measurements at $c_5$ for (a) ECHO and (b) aluminum reflector.}
\label{fig:1D}
\vspace{-3mm}
\end{figure}

Fig.~\ref{fig:1D} shows the measurement results at $c_5$. The x-axis is the path distance which is speed of light multiplied by the path delay, and y-axis is the measurement index for all 341 measurements, which corresponds to a particular scan angle at the transmitter and at the receiver. The plot on the top corresponds to ECHO Metawave reflector, and the plot on the bottom corresponds to the aluminum reflector. The path reflecting from the reflector and arrving at the receiver has a distance of $4.9+3.46=8.36$~m. As $c_5$ is the designed optimum point for ECHO, we observe a strong MPC around that distance. As opposed to the ECHO, the MPC reflecting off the reflector is not the strongest MPC for aluminum reflector at this position. The strongest path and the reflector path are indicated on the plot. For the ECHO measurements, the two coincide. The LOS path around $3.88+1.72=5.6$~m distance is only visible at some antenna positions. As the transmitter and receiver antennas rotate, the power of the MPCs fluctuates.

\begin{figure}[!t]
\centering
\subfloat[]{\includegraphics[width=3.5cm]{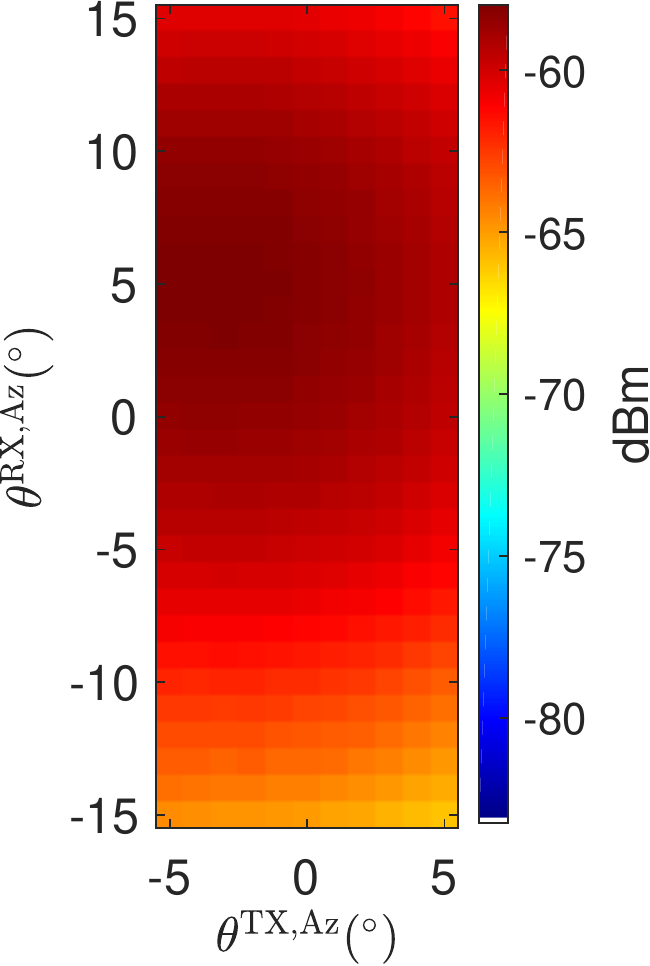}
\label{fig:2Da}}
\hspace{1cm}
\subfloat[]{\includegraphics[width=3.5cm]{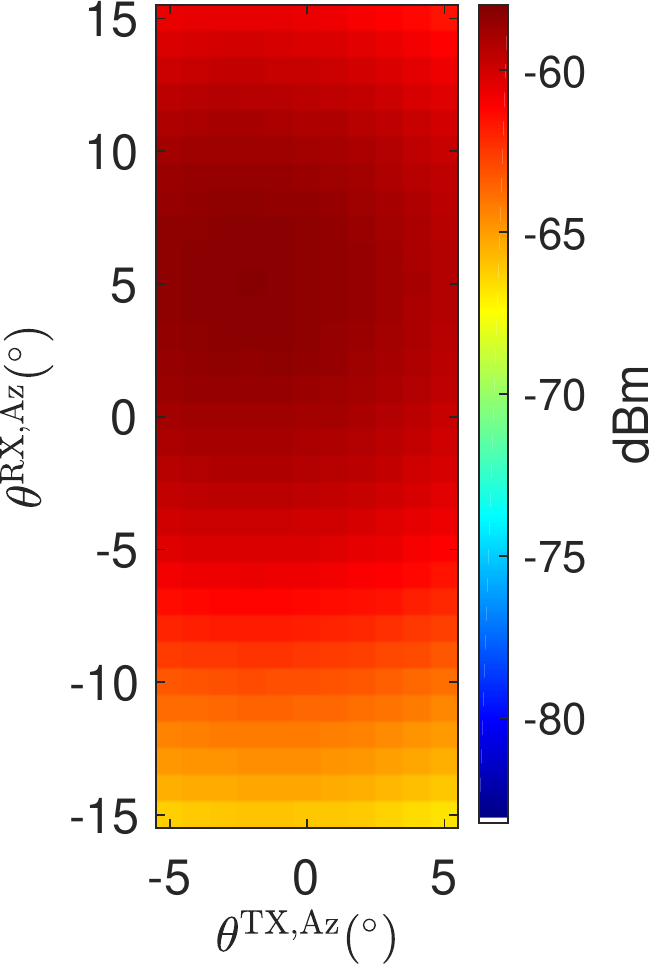}
\label{fig:2Db}}

\subfloat[]{\includegraphics[width=3.5cm]{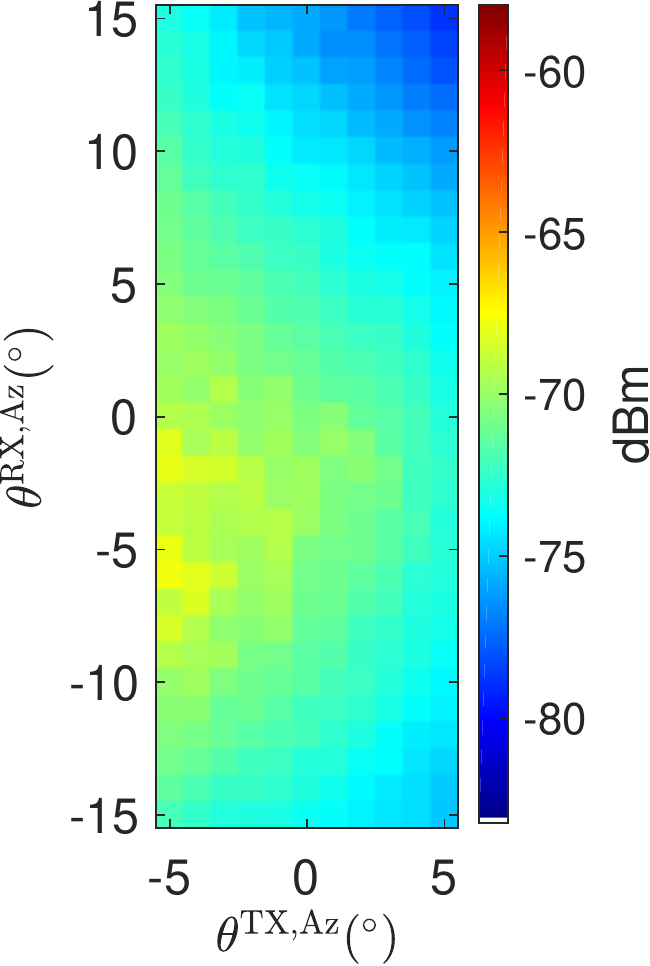}
\label{fig:2Dc}}
\hspace{1cm}
\subfloat[]{\includegraphics[width=3.5cm]{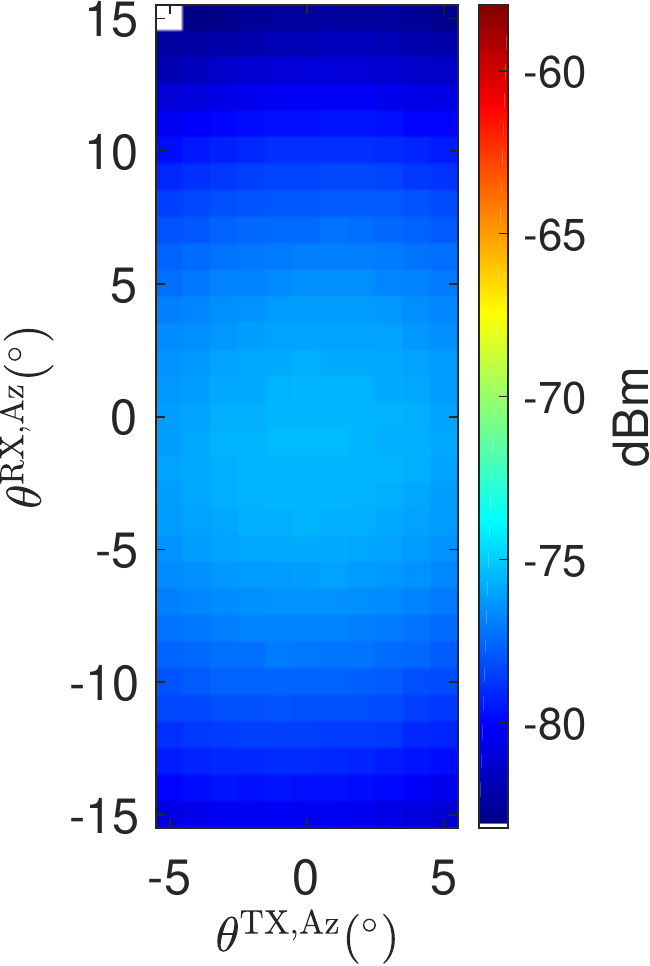}
\label{fig:2Dd}}

\caption{Measurements at indoor $c_5$ position as a function of transmitter and receiver antenna positions. (a) Total power (ECHO), (b) reflector power (ECHO), (c) total power (aluminum), and (d) reflector power (aluminum).}
\label{fig:2D}
\vspace{-3mm}
\end{figure}

In Fig.~\ref{fig:2D}, we plot the power as a function of transmitter and receiver antenna positions for the measurements at $c_5$. The total power is the sum of powers from all the MPCs for that measurement, whereas reflector power corresponds to the path reflecting off the reflector. Note that, at $c_5$ the total power is dominated by the reflector path as Fig.~\ref{fig:2D}(a) is almost identical to Fig.~\ref{fig:2D}(b); however, this is not the case for the aluminum reflector.

\begin{figure}[t!]
\centering
\centerline{\includegraphics[width=0.9\linewidth]{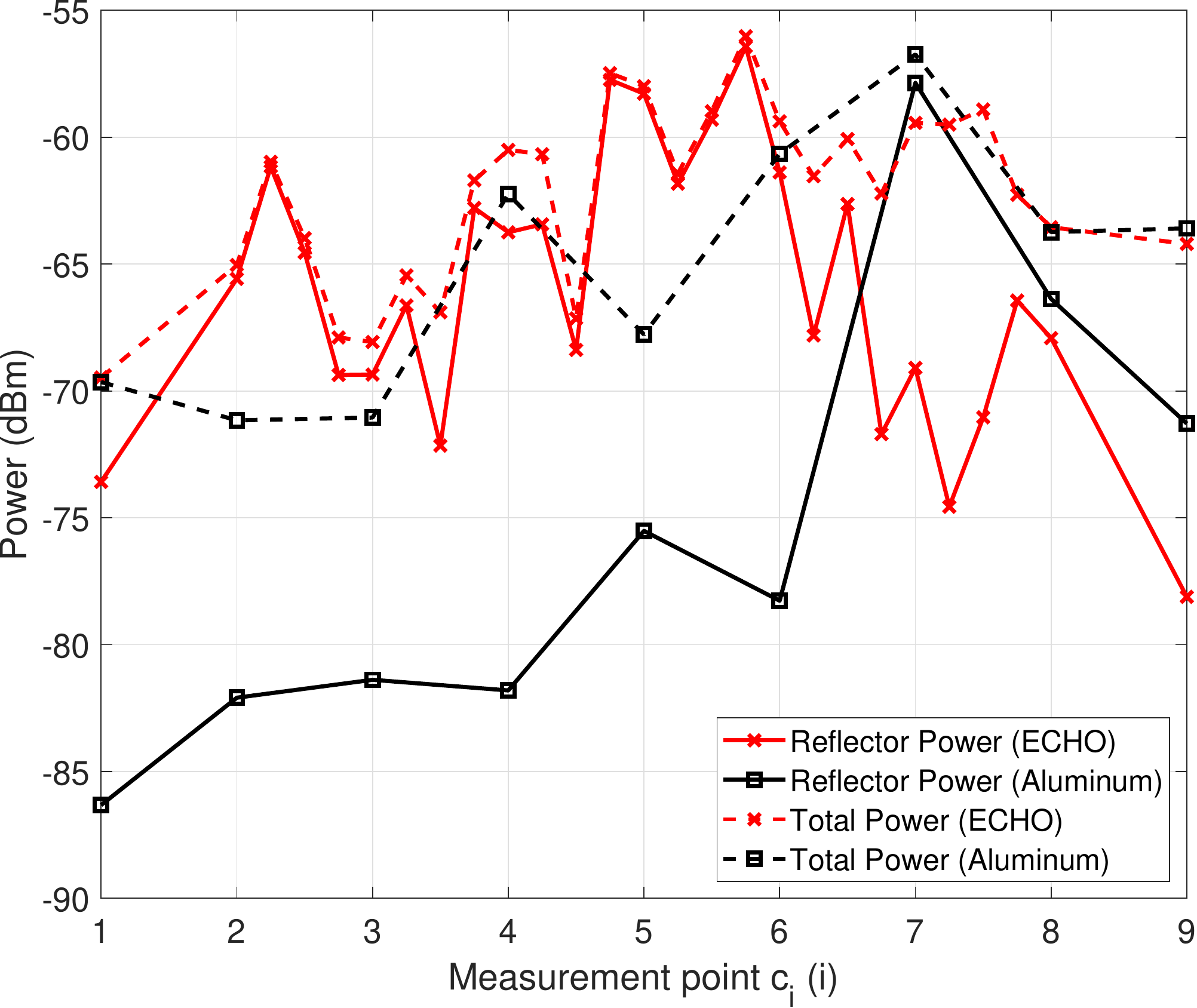}}
\caption{Reflector power and total power results for the indoor measurements.}
\label{fig:indoor}
\end{figure}

The reflector power and total power results from all measurement points $c_1$ to $c_9$ are shown in Fig.~\ref{fig:indoor} for both ECHO Metawave reflector and aluminum reflector. We observe that in case of ECHO Metawave reflector there is significant reflector power around $c_5$ whereas in the case of aluminum reflector we observe more power at later measurement points as expected. For measurement points before $c_6$, reflector power due to ECHO is significantly larger (around 15~dB) than reflector power due to aluminum. However, when the total power is compared, we do not see such an improvement due to ECHO. This is because in an indoor environment there are other reflections that dominate the total received power when the reflector path is not strong enough.   

\begin{figure}[t!]
\centering
\centerline{\includegraphics[width=0.8\linewidth]{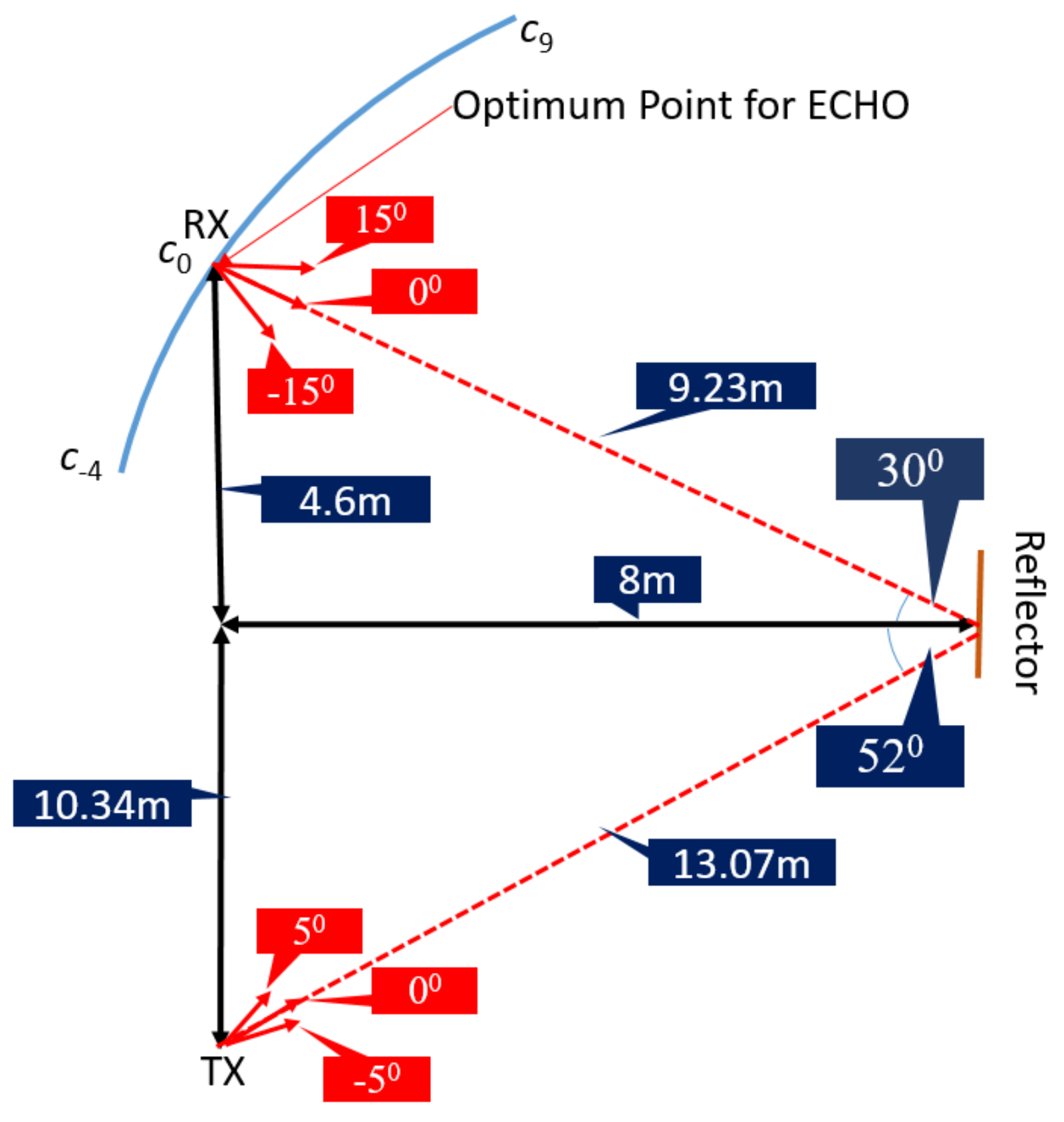}}
\caption{Outdoor setup for ECHO measurements. The measurements are performed along a circle between points $c_{-4}$ and $c_9$. Measurement points are separated by $3.2^{\circ}$.}
\label{fig:SHsetupoutdoor}
\vspace{-3mm}
\end{figure}

We have repeated the ECHO measurements in an outdoor environment, as shown in Fig.~\ref{fig:SHsetupoutdoor} where the distances are longer and the optimum point is labelled as $c_0$. The measurements are performed between $c_{-4}$ and $c_9$ and the measurements points are separated by $3.2^\circ$. 

\begin{figure}[t!]
\centering
\centerline{\includegraphics[width=0.9\linewidth]{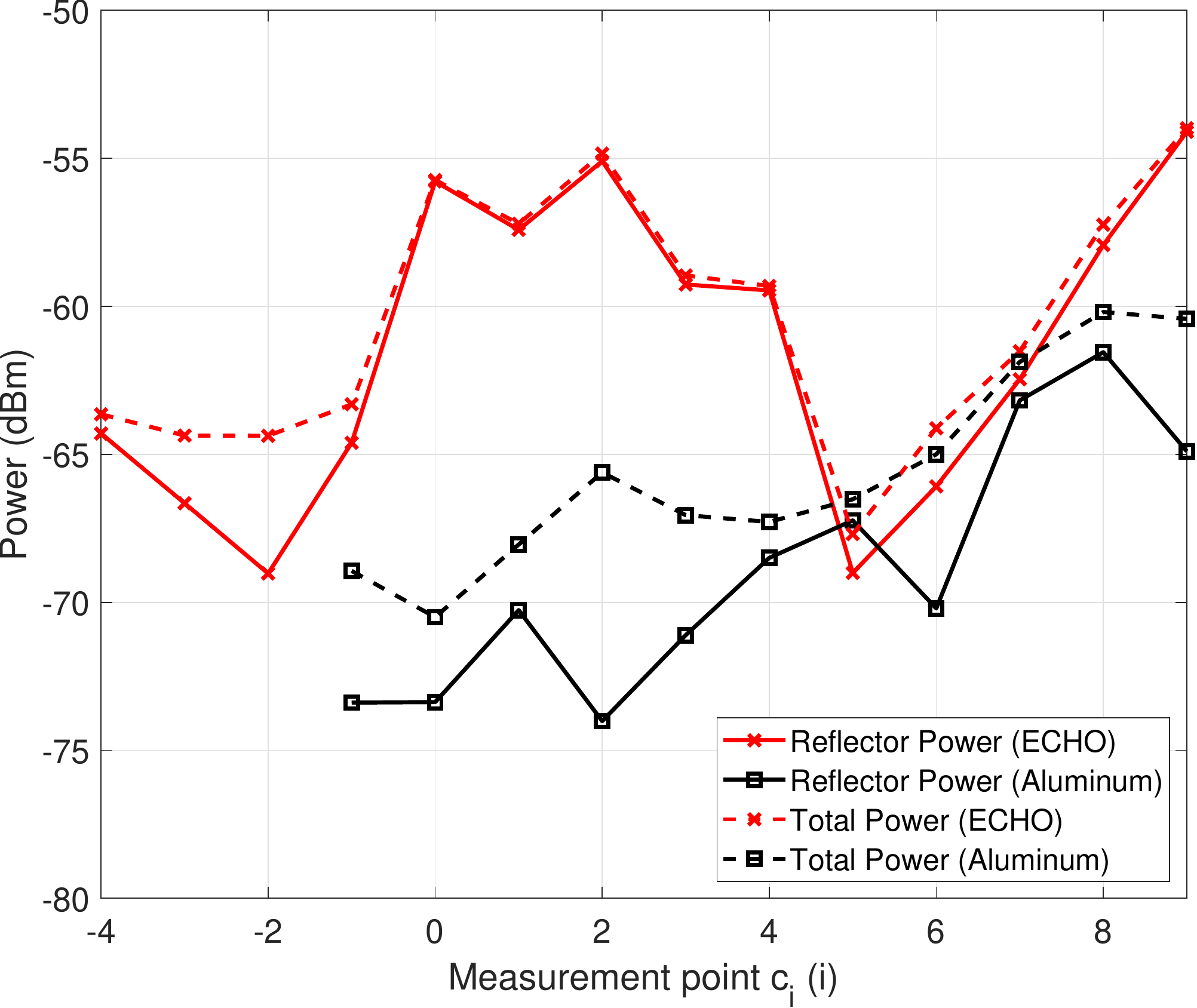}}
\caption{Reflector power and total power results for the outdoor measurements.}
\label{fig:outdoor}
\end{figure}

As shown in Fig.~\ref{fig:outdoor}, ECHO performs significantly better than the aluminum reflector in terms of both total power and the reflector power. This is because the outdoor environment shown in Fig.~\ref{fig:pictures}(b) does not have as many reflections as the indoor environment shown in Fig.~\ref{fig:pictures}(a). Therefore, in the outdoor environment the total power is dominated largely by the reflections from the ECHO or aluminum reflector. 

\section{Coverage Enhancement Using TURBO Active Repeater}
\label{Sec:RepeaterResults}
\begin{figure}[t!]
\centering
\centerline{\includegraphics[trim=7.5cm 6cm 7.5cm 5cm,width=0.9\linewidth]{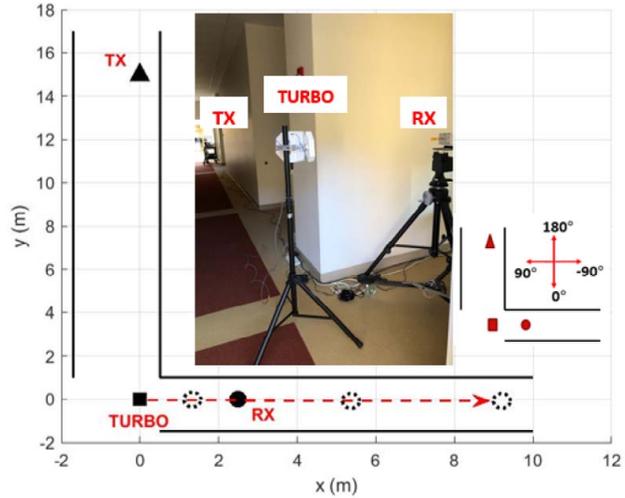}}
\caption{TURBO measurement setup.}
\label{fig:turbo}
\vspace{-3mm}
\end{figure}

An active repeater amplifies a signal received from a direction and re-transmits it in a different direction. We have also performed measurements with TURBO, which is the 28 GHz active repeater solution from Metawave, as shown in Fig.~\ref{fig:turbo}. The measurements are performed in a hallway in the NCSU engineering bulding. Here, the transmitter and the receiver do not have a LOS however the TURBO has LOS with both transmitter and the receiver. The transmitter antenna and the  receiver antenna are rotated $340^\circ$ with $20^\circ$ resolution in the azimuth plane. For each combination of transmitter and receiver antennas the channel PDP is measured and total power for each measurement is calculated. The receiver is located at 4 different positions indicated as circles in the layout in Fig.~\ref{fig:turbo}. 


\begin{figure*}[!t]
\centering
\subfloat[]{\includegraphics[width=0.24\linewidth]{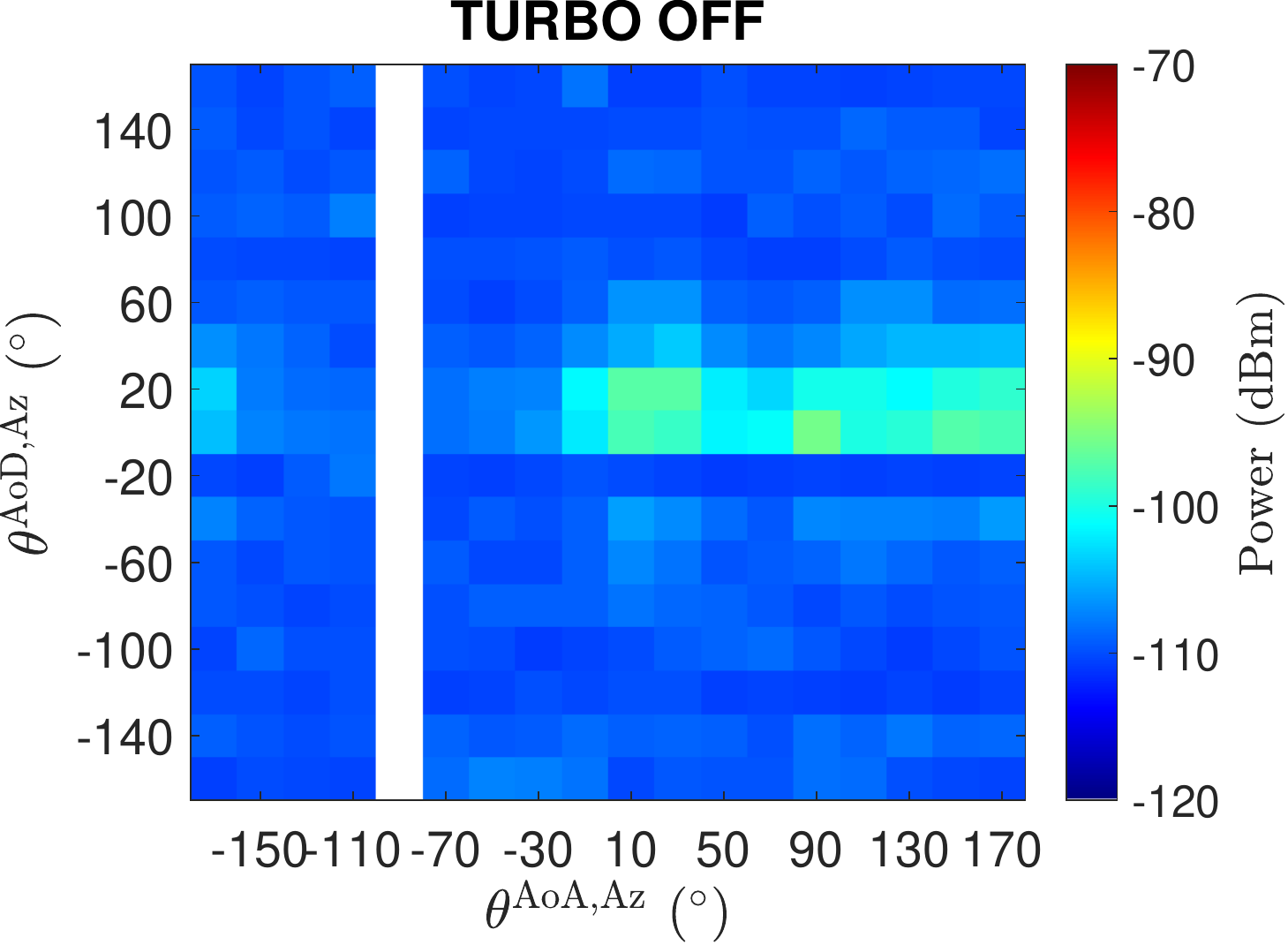}
\label{fig:lasta}}
\subfloat[]{\includegraphics[width=0.24\linewidth]{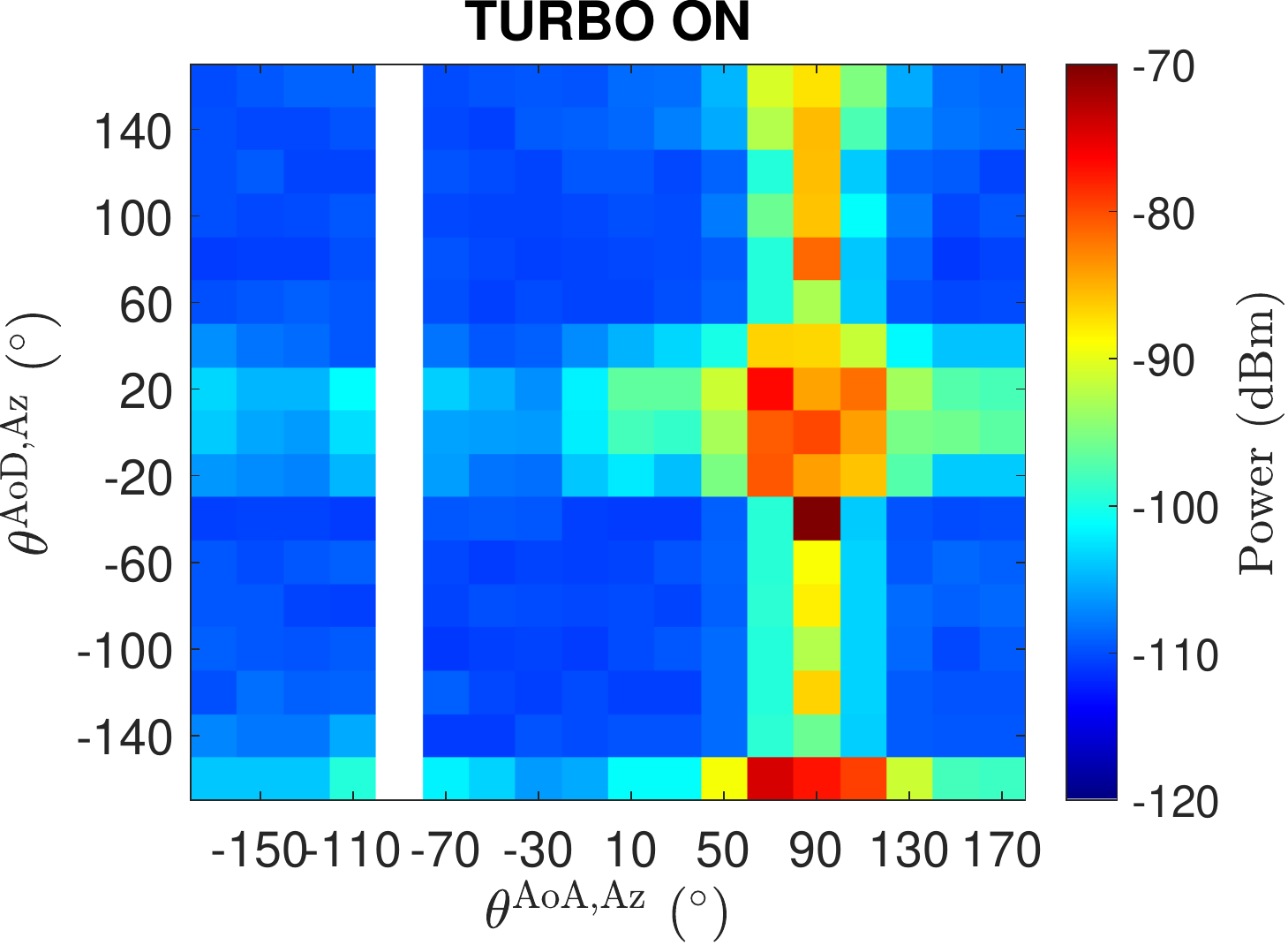}
\label{fig:lastb}}
\subfloat[]{\includegraphics[width=0.24\linewidth]{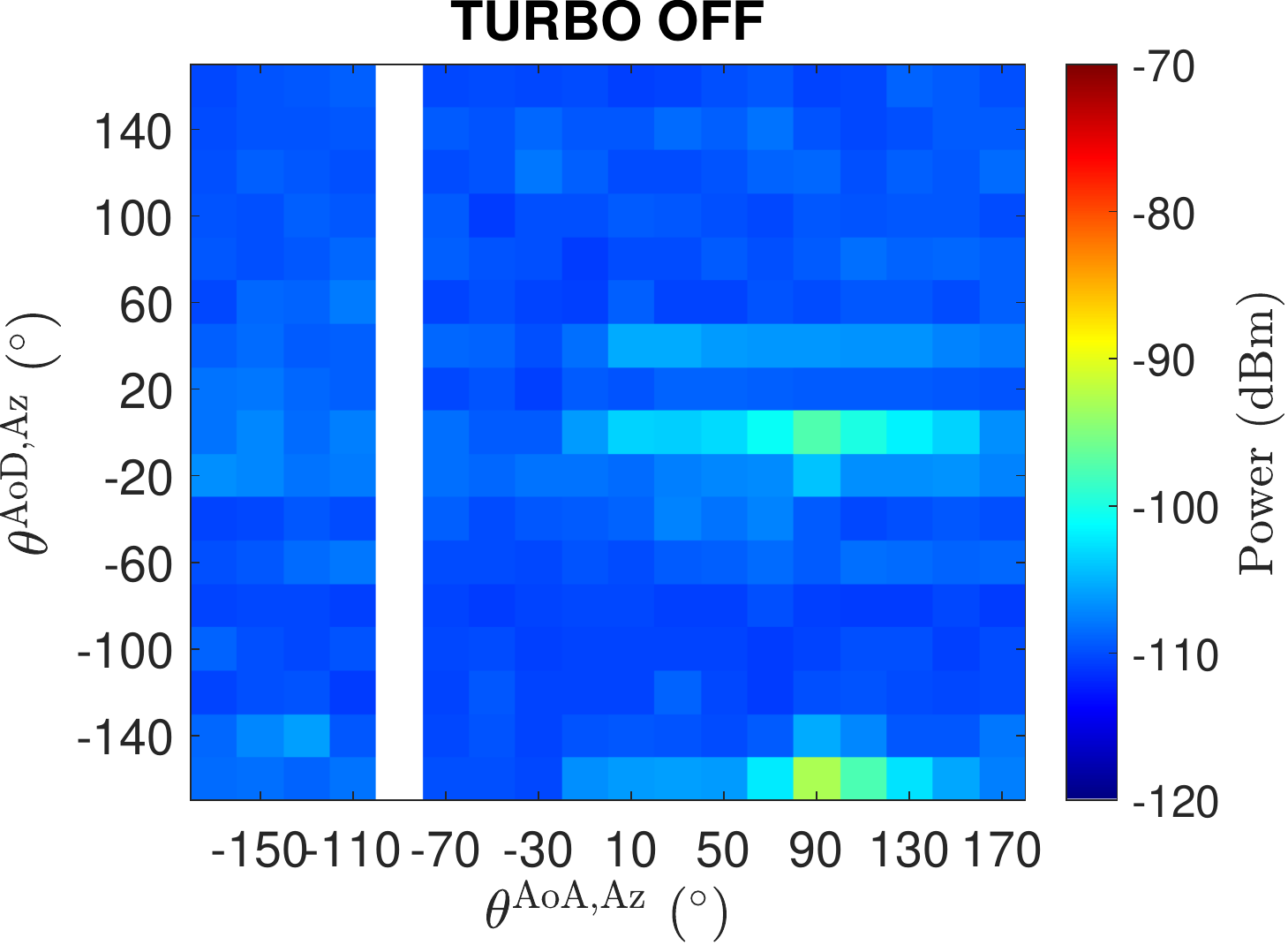}
\label{fig:lastc}}
\subfloat[]{\includegraphics[width=0.24\linewidth]{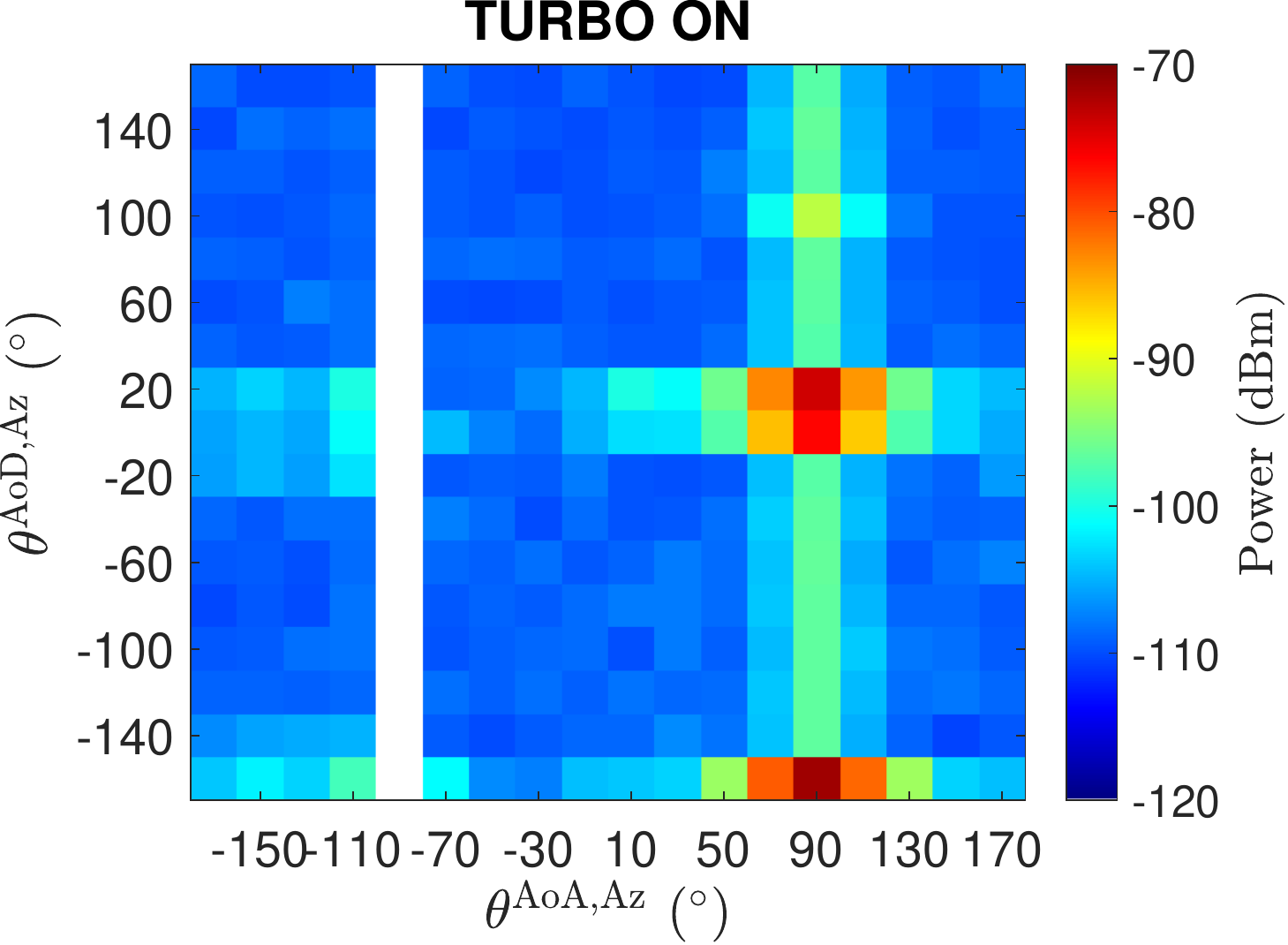}
\label{fig:lastd}}

\subfloat[]{\includegraphics[width=0.24\linewidth]{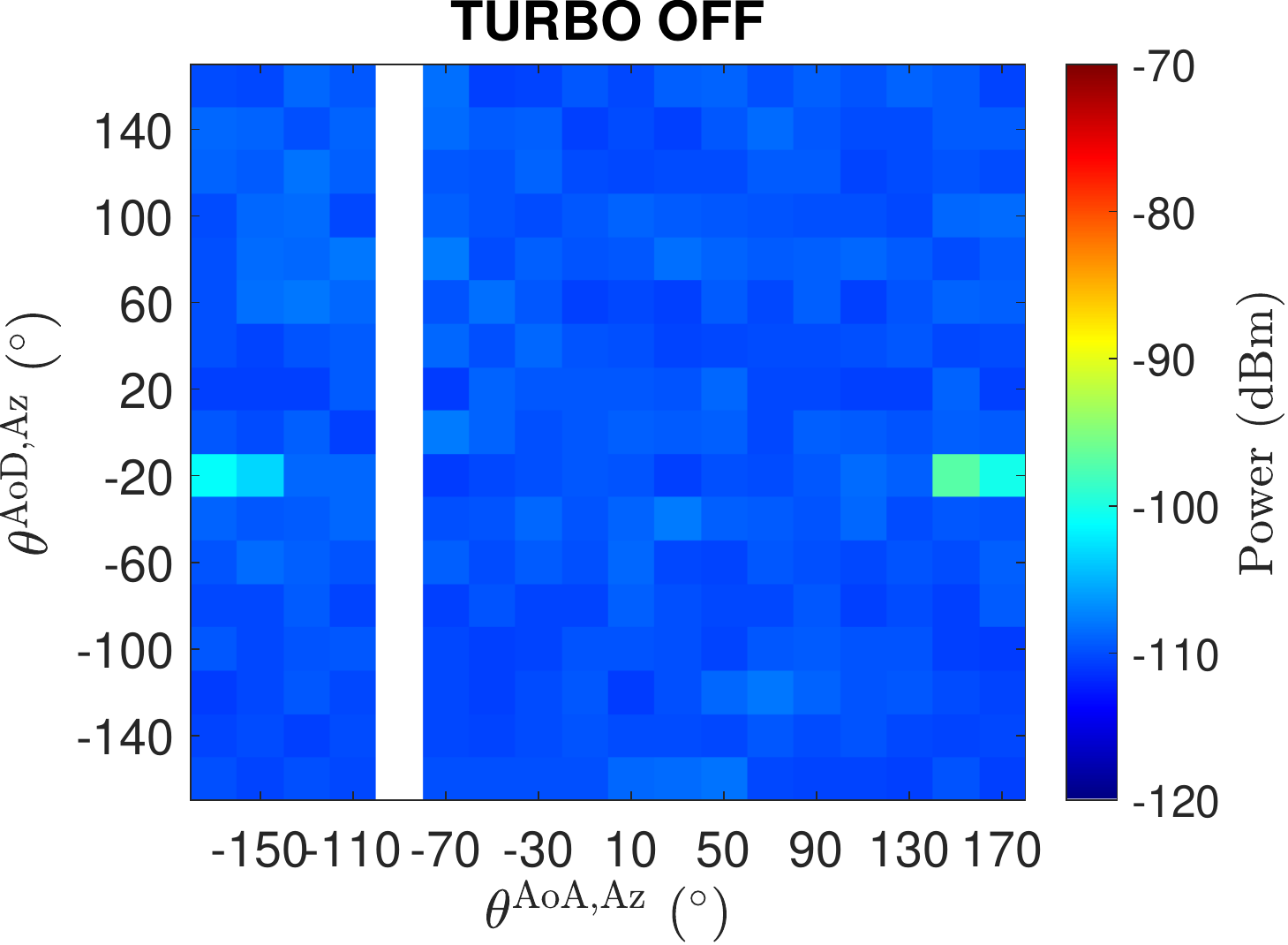}
\label{fig:laste}}
\subfloat[]{\includegraphics[width=0.24\linewidth]{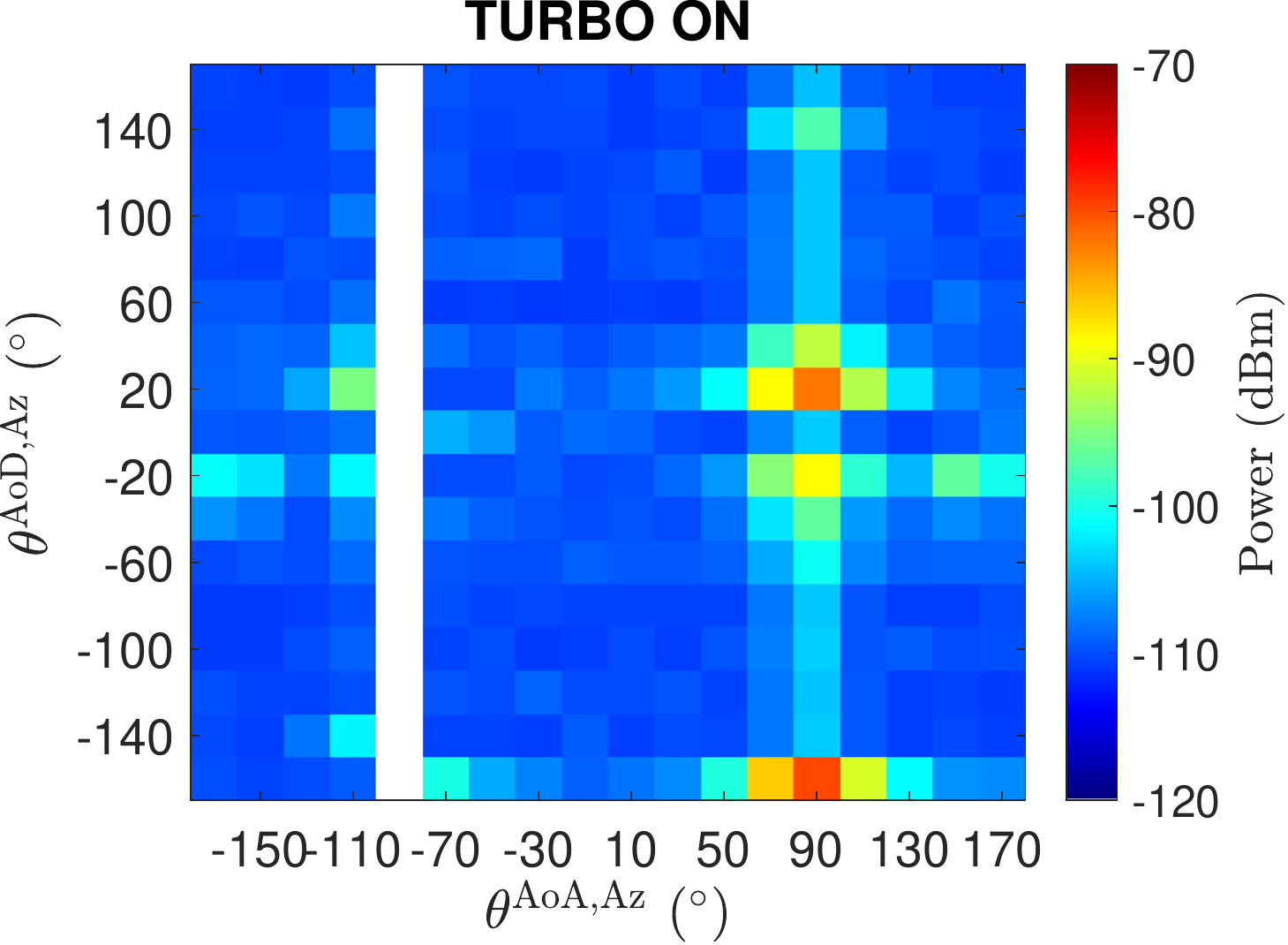}
\label{fig:lastf}}
\subfloat[]{\includegraphics[width=0.24\linewidth]{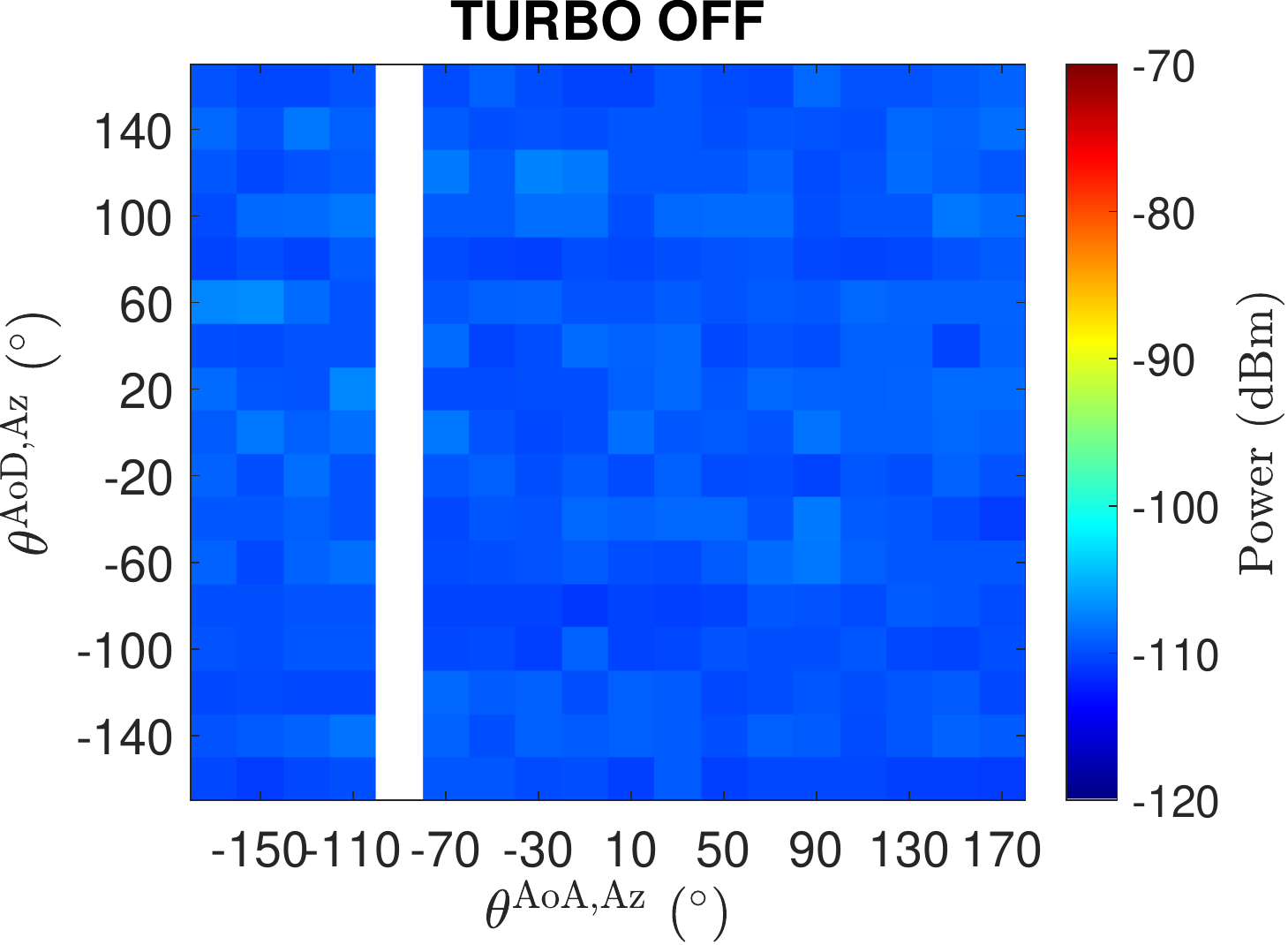}
\label{fig:lastg}}
\subfloat[]{\includegraphics[width=0.24\linewidth]{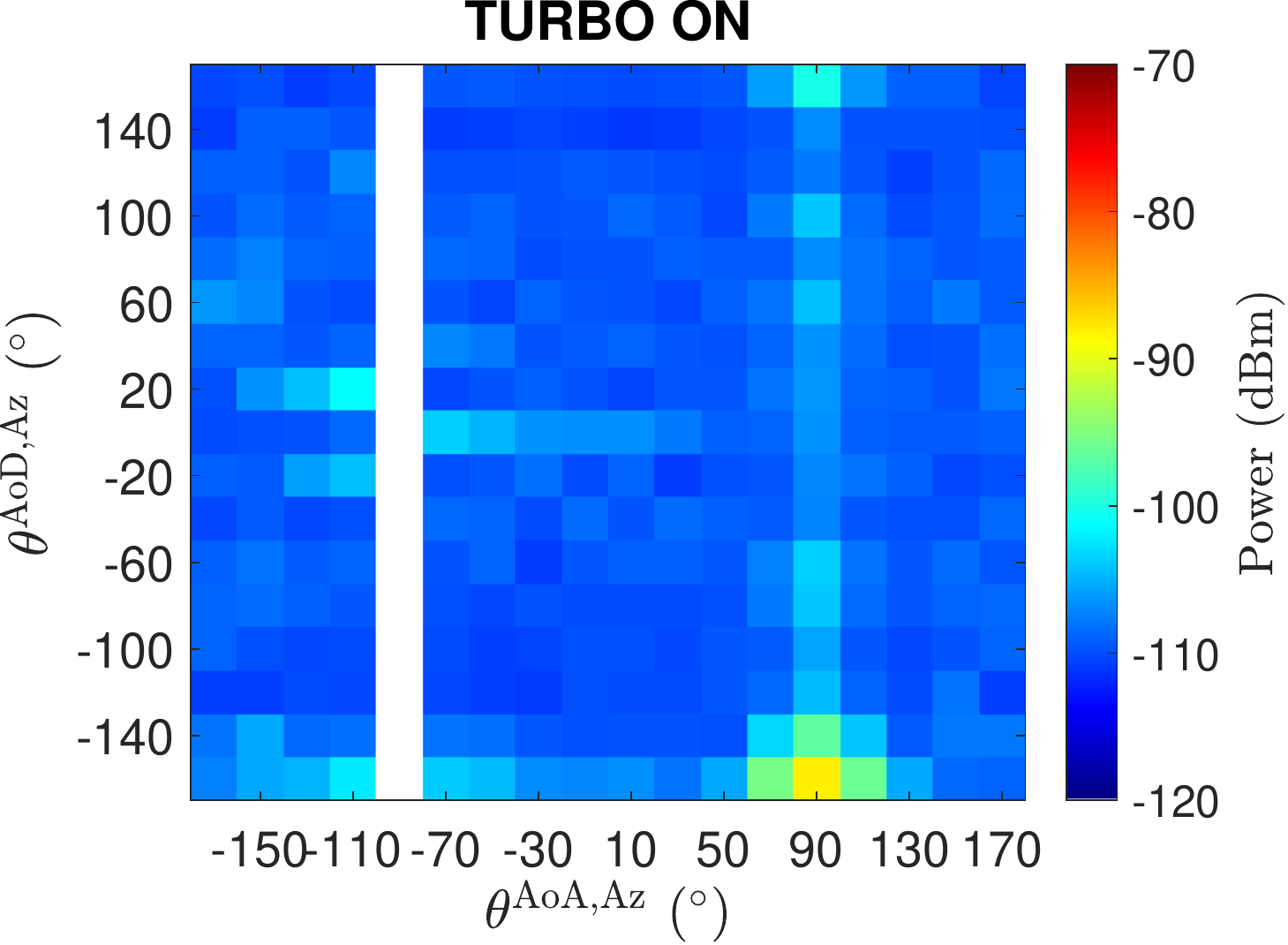}
\label{fig:lasth}}

\caption{Total received power for the setup shown in Fig.~\ref{fig:turbo} when TURBO is ON and OFF. TURBO-RX separation distance: (a)-(b) 1~m, (c)-(d) 2~m, (e)-(f) 5~m, and (g)-(h) 8.5~m.}
\label{fig:TURBOTotPow}
\vspace{-3mm}
\end{figure*}

Fig.~\ref{fig:TURBOTotPow} shows the total received power as a function of the angles of the transmitter and receiver antennas. Eight separate sub-figures correspond to 4 positions and whether TURBO is ON or OFF. It is observed that turning ON the TURBO improves the coverage for all RX positions. The max received powers and gain due to TURBO at each position are tabulated in Table~\ref{Tab:turbo}. TURBO is able to improve the received signal power close to 40 dB at 1 m distance. 

\begin{table}[t]
\renewcommand{\arraystretch}{1.05}
\caption {Measurement Results with TURBO.}
\label{Tab:turbo}
\centering
{\begin{tabular}{|c|c|c|c|}
\hline
 & TURBO OFF &  TURBO ON &  \\ \hline
RX-TURBO & Max Power & Max Power & Gain due to  \\ 
distance (m) & (dBm) & (dBm) & TURBO (dB) \\ \hline \hline
1 & -95.49 & -56.57 & 39.82 \\ 
2 & -92.80 & -71.56 & 21.24 \\
5 & -97.03 & -79.77 & 17.26 \\
8.5 & -106.88 & -88.02& 18.86\\\hline
\end{tabular}}
\end{table}

\section{Conclusions}
\label{Sec:Conclusion}
In this paper, we compared 28 GHz mmWave PADP measurement results with horn antennas and phased array antennas. We showed that both antennas can be used to extract angular profile of the mmWave channel. We have also introduced ECHO passive reflector and TURBO active repeater solutions from Metawave, Inc. to improve coverage and extend the range of mmWave signal. 

\section*{Acknowledgment}
The authors would like to thank Ender Ozturk, Kairui Du, Mrugen Deshmukh, Sultan Almutairi, and Bill Mwaniki from NC State University for their help with measurements, Metawave Corporation, Inc. for providing ECHO and TURBO, and Nadisanka Rupasinghe, Haralabos  Papadopoulos, and Fujio Watanabe,  from DOCOMO Innovations, Inc. for their suggestions for the measurements. 

\bibliography{IEEEabrv,references}
\bibliographystyle{IEEEtran}

\end{document}